\documentclass[11pt,showpacs,showkeys,preprintnumbers]{revtex4-2}

\usepackage{graphicx}   
\usepackage{color}        
\usepackage{braket}
\usepackage{amsmath}
\usepackage{amssymb}
\usepackage{mathrsfs} 
\usepackage{titlesec}
\usepackage{makecell}
\usepackage{natbib,hyperref}
\usepackage{tabularray}
\usepackage{multirow}
\usepackage[mathscr]{euscript}

\def \be  {\begin{equation}}
\def \ee  {\end{equation}}
\def \bea {\begin{eqnarray}}
\def \eea {\end{eqnarray}}

\usepackage{mathrsfs}  

\definecolor{ao-english}{rgb}{0.0, 0.5, 0.0}
\definecolor{cadmiumgreen}{rgb}{0.0, 0.42, 0.24}

\newcommand{\nn}{\nonumber}

\begin{document}

\preprint{ECTP-2025-04}
\preprint{WLCAPP-2025-04}
\hspace{0.05cm}

\title{Temperature Dependence of the Masses of Various Meson States: A Comparative Study in SU($3$) and SU($4$) extended Linear-Sigma Model}
\email{tawfik@itp.uni-frankfurt.de; 400778@iu.edu.sa, atawfik@bnl.gov}

\author{A.~Friesen$^{1}$, Yu. Kalinovsky$^{1}$, S.~O.~Allehabi$^{2}$, N.~M.~Rfeek$^{3}$, A.~A.~Alshehri$^{4}$, A.~Tawfik$^{2,5}$}

\affiliation{$^{1}$Joint Institute for Nuclear Research (JINR), 141980 Dubna, Russia Federation}  
\affiliation{$^2$Department of Physics, Faculty of Science, Islamic University of Madinah, 42351 Madinah, Saudi Arabia} 
\affiliation{$^3$Physics Department, Faculty of Science, Assiut University, 71515 Assiut, Egypt} 
\affiliation{$^4$Department of Science and Technology, University of Hafr Al Batin (UHB), University College at Nairiyah, Nairiyah 31981, Saudi Arabia} 
\affiliation{$^5$Basic Science Department, Faculty of Engineering, Ahram Canadian University (ACU), 12556 Giza, Egypt} 

\begin{abstract}
In the extended Linear-Sigma Model (eLSM), the chiral phase structure of meson states, including pseudoscalars ($J^{pc}=0^{-+}$), scalars ($J^{pc}=0^{++}$), vectors ($J^{pc}=1^{--}$), and axial-vectors ($J^{pc}=1^{++}$), is investigated with the mean-field approximation. A systematic comparison between SU($3$) and SU($4$) configurations is provided. It has been found that the estimations of meson masses derived from SU($4$) eLSM are more congruent with experimental values than those derived from SU($3$) eLSM. Consequently, we conclude that an increase in quark degrees of freedom significantly enhances the accuracy of meson mass simulations. We investigate the effect of temperature on the masses of various meson states calculated in the SU($3$) and SU($4$) eLSM. After establishing all the fitting parameters, the temperature dependence of meson masses shows that although various meson states exhibit unique patterns in their mass changes with temperature, they all seem to share a similar range of dissolution temperatures. This means that the critical temperature that marks the phase transition from hadrons to quarks appears to vary slightly depending on the meson states. In this regard, we find that the quarkonium states, formed by a quark and its antiquark, are largely unaffected by variations in the temperature.

\end{abstract}


\maketitle

\section{Introduction}
\label{sec:intrd}

To analyze different properties of QCD, several low-energy effective QCD-like models have been established, thereby providing insights into the phase structure at finite temperatures, densities, and in the presence of magnetic and electric fields \cite{Tawfik:2014uka,Tawfik:2016lih,Tawfik:2016gye,Tawfik:2016edq}. In the limit of vanishing quark masses, chiral symmetry emerges as a fundamental symmetry in QCD \cite{Pagels:1978ba,Gasiorowicz:1969kn,Becchi:1980vz,Trueman:1979en}. On the other hand, in the limit of finite quark masses, chiral symmetry becomes spontaneously broken. In this limit, the chiral condensate represents the associated order parameter. With small quark masses, the explicit breaking of chiral symmetry for the up and down quarks \cite{Osipov:2013fka} causes the pseudo-Goldstone bosons to acquire finite masses, such as the light pseudoscalar mesons, which include pions \cite{Freese:1990rb,Foot:2007as}. For heavier quark flavors, including strange and charm quarks, the explicit breaking intensifies, resulting in the emergence of both hidden and open charmed mesons like $D$ \cite{Yamada:2001mv} and $\chi_{c0}$ meson \cite{Wilson:2023anv}, respectively. Under extreme conditions of high temperature and/or density, the in-medium chiral condensate seems to vanish, demonstrating the restoration of chiral symmetry and the emerging of meson state degeneracy \cite{Jafarov:2003pe,Matsuo:2008cm}.

The QCD phase structure can be explored using low-energy QCD methods, such as Dyson--Schwinger equations \cite{Zhang:2019gva,Fischer:2018sdj,Ayala:2011vs} and chiral perturbation theory \cite{Espriu:2020dge}, etc. Different statistical thermal models are also employed in examining QCD phase structure. This includes, for instance the Hadron Resonance Gas (HRG) \cite{Karsch:2003vd,Karsch:2003zq} Nambu-Jona-Lasinio (NJL) \cite{Zhao:2020xob,Buballa:2003qv}, and Linear-Sigma Model (LSM) \cite{Tawfik:2014gga,Tawfik:2016gye,Tawfik:2019rdd,Tawfik:2019rdd}. In effective QCD-like models, such as LSM
\cite{Tawfik:2014gga,Tawfik:2016gye,Tawfik:2019rdd,Tawfik:2019rdd}, pion mesons can be contracted with two quark flavors, specifically in an SU(2) configuration
\cite{Gallas:2009qp,Parganlija:2010fz,Janowski:2011gt}. This framework also facilitates a systematic examination of light quark condensates \cite{Narison:1988xi}. The SU($3$) configuration enables the construction of nonet meson masses
\cite{Tawfik:2016gye,Tawfik:2019rdd}, the study of the QCD phase transition
\cite{tHooft:1976rip,Parganlija:2012fy}, and the investigation of strange condensate
\cite{Tawfik:2016gye,Tawfik:2014uka}. The SU(4) configuration evidently incorporates the charm quark, allowing for the analysis of the corresponding condensate. A systematic investigation of charmed meson masses was analytically derived and presented in Ref.
\cite{Tawfik:2025hhr}. This manuscript is devoted to exploring the in-medium modifications of various charmed meson masses. A systematic analysis comparing the temperature dependence of various meson masses of SU(3) and SU(4) is also conducted.

Recent advancements in experimental high-energy particle physics, particularly at the Large Hadron Collider (LHC) at CERN and the Relativistic Heavy Ion Collider (RHIC) at Brookhaven National Laboratory (BNL), especially the Beam Energy Scan program of the STAR experiment \cite{Vertesi:2009wf}, have brought considerable prominence among researchers to the in-medium modifications of various hadron states \cite{Cassing:2002pi,Dainese:2003wq,Tawfik:2006yq}. The analysis of in-medium modifications of various hadron properties is anticipated in future facilities including the Facility for Antiproton and Ion Research (FAIR), at GSI, Darmstadt-Germany \cite{Friese:1999qm,Tolos:2005ft,Finogeev:2018ing} and  Nuclotron-based Ion Collider fAcility (NICA) at JINR, Dubna-Russia \cite{Parfenov:2021isv}. 

In-medium modifications of charmed mesons are particularly interesting because they serve as probes for the hot and dense medium created during heavy-ion collisions. However, modeling the charm sector using effective models like the eLSM comes with its own set of challenges, especially due to the necessity of an extra mass term $-2 Tr[\epsilon \Phi^{\dagger} \Phi]$ \cite{Giacosa:2012cq}, along with the complex interactions between light and heavy flavor condensates. We would like to emphasize that our calculations incorporate the $U(1)_A$ anomaly. This is $c (\rm{det}\Phi+\rm{det}{\Phi^\dagger})$-term. Moreover, the influence of the anomaly on the pseudo-critical temperatures is also considered. More comprehensive information about the anomaly terms in the Lagrangian concerning phase transitions can be found in the Ref.\cite{Pisarski:2024esv,Giacosa:2024orp}.

To integrate the finite-temperature behavior of the condensates into the model, we apply the mean-field approximation and account for finite-temperature effects via the grand potential, which includes contributions from quark-antiquark pairs and the Polyakov loop. We define a temperature-dependent mass parameter $m_0^2$ as $m^2_0 (1-T^2/T_c^2)$. This change is motivated by the fact that in the linear sigma model, the critical temperature scales as $N_c^{1/2}$ in the limit of $N_c \rightarrow \infty$, which is inconsistent with the $N_c$-independent behavior found in the NJL model \cite{Meyer:2001zp}. The inclusion of this temperature-dependent factor appears to restore the expected large-$N_c$ scaling behavior  \cite{Heinz:2011vs,Heinz:2011xq}.

The quadratic dependence on temperature $T$ is motivated by the low-temperature results of chiral perturbation theory, where the chiral condensate decreases in proportion to $T^2$ due to thermal fluctuations of Goldstone bosons \cite{Gasser:1986vb}. In this regard, the factor introduced seems to model the chiral dynamics of the system. Nonetheless, this factor alone appears inadequate to describe the chiral phase transition or to identify the critical temperature. On the contrary, the factor rather illustrates the gradual decrease of the chiral condensates as temperature increases. The gluonic sector, which is responsible for confinement dynamics, is effectively incorporated through the Polyakov loop. By integrating both elements, we can simultaneously recover the correct large-$N_c$ behavior and enforce the chiral effects and back-reaction of quarks into the model.

The effectiveness of the SU($3$) eLSM in examining light meson states and their chiral phase structure has been established
\cite{Tawfik:2014gga,Tawfik:2019rdd,Kovacs:2016juc}. Nonetheless, there exists a gap in the systematic and comparative analysis of meson masses in SU($4$) configurations at finite temperature. The present study aims to present a comprehensive description of various meson states and their temperature dependence, all within the eLSM framework for both SU($3$) and SU($4$) flavor symmetries.

This manuscript is structured as follows. In Section \ref{sec:Lagrangian}, the extended Linear Sigma Model (eLSM) is reviewed. The configurations for SU($3$) and SU($4$) are discussed in Sections \ref{sec:su3} and \ref{sec:su4}, respectively. Section \ref{sec:finiteT} provides a detailed explanation of the finite temperature formalism for the masses of various meson states. The parametrization along with the numerical results is examined in Section \ref{sec:rslts}. The final conclusions are presented in Section \ref{sec:cnls}.

\section{Extended Linear Sigma Model} 
\label{sec:Lagrangian}

The LSM \cite{Mignaco:1971yr,Contreras:1989gi,Bhattacharyya:1995xt,Roh:1996ek,Delbourgo:1998kg,Lenaghan:2000ey} provides a remarkably accurate description of various meson states \cite{Tawfik:2014gga,Tawfik:2019rdd,Tawfik:2025hhr}. As the degrees of freedom related to the quark flavors $N_f$, increase, so does the number of meson states that can be created. The SU($3$) meson states were derived and introduced in Refs. \cite{Tawfik:2014gga,Tawfik:2019rdd}. The analytical derivation of SU($4$) meson states is introduced in Refs. \cite{Ahmadov:2023mmy, Tawfik:2025hhr}. 

The Lagrangian for the mesonic sector, which contains scalar, pseudoscalar, vector, and axial-vector mesons, together with their interactions and anomalies, is built as follows \cite{Tawfik:2025hhr}: 
\begin{equation}
    \mathcal{L}=\mathcal{L}_{SP}+ \mathcal{L}_{VA}+ \mathcal {L}_{Int}+ \mathcal{L}_{U(1)_A},
    \label{eq:Lagrangian}
\end{equation}
\begin{eqnarray}
\mathcal{L}_{SP}&=&\mathrm{Tr}\left[(D^{\mu}\Phi)^{\dagger}\,(D^{\mu}\Phi)-m^2
\Phi^{\dagger} \Phi\right]-\lambda_1 [\mathrm{Tr}(\Phi^{\dagger} \Phi)]^2
-\lambda_2 \mathrm{Tr}(\Phi^{\dagger}
\Phi)^2 + \mathrm{Tr}[H(\Phi+\Phi^{\dagger})], \label{eq:scalar_nonets} \\
  \mathcal{L}_{AV}&=&-\frac{1}{4}\mathop{\mathrm{Tr}}(L_{\mu\nu}^{2}+R_{\mu\nu}^{2}
)+\mathop{\mathrm{Tr}}\left[  \left( \frac{m_{1}^{2}}{2}+\Delta\right)  (L_{\mu}^{2}+R_{\mu}^{2}    )\right] \nonumber \\
&+&i \frac{g_{2}}{2} (\mathop{\mathrm{Tr}}\{L_{\mu\nu}[L^{\mu},L^{\nu}]\}+\mathop{\mathrm{Tr}}\{R_{\mu\nu}[R^{\mu},R^{\nu}]\}){\nonumber}\\
&+& g_{3}[\mathop{\mathrm{Tr}}(L_{\mu}L_{\nu}L^{\mu}L^{\nu}
)+\mathop{\mathrm{Tr}}(R_{\mu}R_{\nu}R^{\mu}R^{\nu})]+g_{4}
[\mathop{\mathrm{Tr}}\left(  L_{\mu}L^{\mu}L_{\nu}L^{\nu}\right)
+\mathop{\mathrm{Tr}}\left(  R_{\mu}R^{\mu}R_{\nu}R^{\nu}\right)
]{\nonumber}\\
&+&g_{5}\mathop{\mathrm{Tr}}\left(  L_{\mu}L^{\mu}\right)
\,\mathop{\mathrm{Tr}}\left(  R_{\nu}R^{\nu}\right)  +g_{6}
[\mathop{\mathrm{Tr}}(L_{\mu}L^{\mu})\,\mathop{\mathrm{Tr}}(L_{\nu}L^{\nu
})+\mathop{\mathrm{Tr}}(R_{\mu}R^{\mu})\,\mathop{\mathrm{Tr}}(R_{\nu}R^{\nu
})],\label{eq:vector_nonets}\\
 \mathcal{L}_{Int}&=&\frac{h_{1}}{2}\mathop{\mathrm{Tr}}(\Phi^{\dagger}\Phi
)\mathop{\mathrm{Tr}}(L_{\mu}^{2}+R_{\mu}^{2})+h_{2}%
\mathop{\mathrm{Tr}}[\vert L_{\mu}\Phi \vert ^{2}+\vert \Phi R_{\mu} \vert ^{2}]+2h_{3}%
\mathop{\mathrm{Tr}}(L_{\mu}\Phi R^{\mu}\Phi^{\dagger}),\label{eq:INT}
\\ 
\mathcal{L}_{U(1)_A}&=&c[\mathrm{Det}(\Phi)+\mathrm{Det}(\Phi^{\dagger})].
 \label{eq:LagrSum}
\end{eqnarray}
The {\it complex} matrices for scalars $\sigma_{a}$, i.e., $J^{PC}=0^{++}$, pseudoscalars  $\pi _{a}$, i.e. $J^{PC}=0^{-+}$, vectors $V_{a}^{\mu}$, i.e., $J^{PC}=1^{--}$ and axial-vectors $A_{a}^{\mu}$, i.e., $J^{PC}=1^{++}$ meson states can be constructed as
\begin{equation}
\label{fieldmatrix}
\Phi 	 = \sum_{a=0}^{N_{f}^{2} -1} T_{a}(\sigma _{a}+ i \pi _{a}), \qquad  \qquad
L^{\mu}  = \sum_{a=0} ^{N_{f}^{2} -1} \, T_{a}\, (V_{a}^{\mu}+A_{a}^{\mu}), \qquad  \qquad
R^{\mu}  = \sum_{a=0}^{N_{f}^{2} -1}\, T_{a}\, (V_{a}^{\mu}-A_{a}^{\mu}).
\end{equation}
The various generators are defined according to the number of quark flavors $N_f$, while $T_{a}$ are the corresponding generators of $U(N_f)$ can be expressed as $T_{a}=\hat{\lambda}_{a}/2$, with $a=0\dots (N_f^2-1)$ and $\hat{\lambda}$ are the Gell--Mann matrices.  

The covariant derivative
\begin{equation}
    D^\mu \Phi \equiv \partial^\mu \Phi-i\,g_1 (L^\mu \Phi - \Phi R^\mu) - i eA^\mu[T_3, \Phi],
    \label{eq:covarD}
\end{equation}
is to be associated with the degrees of freedom for (pseudo-)scalar and (axial-)vector and couples them through the coupling constant $g_1$.  
\begin{eqnarray}
    L^{\mu \nu} &\equiv & \partial^\mu L^\nu - i eA^\mu[T_3, L^\nu] -\lbrace  \partial^\nu L^\mu -  i eA^\nu[T_3, L^\mu]\rbrace,\\
    R^{\mu \nu} &\equiv & \partial^\mu R^\nu - i eA^\mu[T_3, R^\nu] -\lbrace  \partial^\nu R^\mu -  i eA^\nu[T_3,R  ^\mu]\rbrace,
\end{eqnarray}
where $A^\mu = g A_\mu^a\lambda^a/2$ is the electromagnetic field. The constant g is the Yukawa coupling, which is fixed from the non-strange constituent quark mass as $g = 2 m_q/\bar{\sigma}_x$. The field $\Phi$ is expressed as
\bea
\Phi &=& \sum_{a=0}^{N_f^2-1} T_a\left(\sigma_a+i\pi_a\right).
\eea
Expressions for $T_a \sigma_a$ and $T_a \pi_a$ can be found in Refs. \cite{Ahmadov:2023mmy, Tawfik:2025hhr}. According to Refs. \cite{Ahmadov:2023mmy, Tawfik:2025hhr}, the global chiral invariance of the Lagrangian for $N_f=4$ is identical to that of $N_f=3$. However, for $N_f=4$, the mass term $\mathcal{L}_{\mathtt{emass}} = -2 \rm{Tr}[\epsilon \Phi^\dagger\Phi]$ must be included \cite{Giacosa:2012cq}. The origin of this term can be understood from the equivalence between the symmetry-breaking Hamiltonians of the $SU(4)\times SU(4)$ and $SU(3) \times SU(3)$ group  \cite{Mott:1974vu,Myers:2007vc}.  

Starting with the SU($3$) configuration of eLSM, the following section goes over the essential characteristics.

\subsection{SU(3) configuration}
\label{sec:su3}

The tree-level mesonic potential for the scalar-pseudoscalar states is presented in the Appendix \ref{sec:App1}, Eq. (\ref{eq:GenMesPot}). For the SU($3$), this is reduced to 
\begin{eqnarray}\label{eq:MesPotSU3}
U({\sigma}) = 
\frac{m^2}{2}{\sigma}_a^2
- {\cal G}_{abc}{\sigma}_{c}{\sigma}_a {\sigma}_b 
 + \frac{1}{3} {\cal F}_{abcd}{\sigma}_a{\sigma}_b{\sigma}_c {\sigma}_d 
- h_a{\sigma}_a, 
\end{eqnarray}
where the coefficients ${\cal G}_{ab}$, ${\cal G}_{abc}$, ${\cal G}_{abcd}$, ${\cal F}_{abcd}$, ${\cal H}_{abcd}$ are detailed in the Appendix \ref{sec:App1}. 
 
It is more convenient to undertake the subsequent analysis in terms of the pure non-strange and strange fields, which are obtained respectively from the following transformation: 
\begin{eqnarray}
    \left( \begin{array}{c}
 \sigma_x  \\
\sigma_y   \\
\end{array}\right) =  \frac{1}{\sqrt{3}}\left(
    \begin{array}{cc}
 \sqrt{2}     &1    \\
1    &  -\sqrt{2}   \\  
\end{array}\right)
\left( \begin{array}{c}
 \sigma_0  \\
\sigma_8   \\
\end{array}\right).
\label{eq:SU3transition}
\end{eqnarray}
The explicit chiral symmetry breaking parameters in the pseudo-scalar sector, i.e., the expression $\rm{Tr}[H(\Phi +\Phi^\dagger)]$ in Eq. (\ref{eq:Lagrangian}), allowing the transformation of $h_0$ and $h_8$ using the same transformation basis
\begin{eqnarray}
    H_{SU(3)} =T_0 h_0+ T_8 h_8\equiv \frac{1}{2}\left(
    \begin{array}{ccc}
 h_{x}     &0          & 0  \\
0    &  h_{x}           & 0   \\
0 & 0 & \sqrt{2} h_{y}  
\end{array}\right).
\end{eqnarray}
Then the tree-level mesonic potential for the SU(3) configuration reads
\begin{eqnarray}
    U(\sigma_x,\sigma_y) &=& \frac{m^2}{2}\left(\sigma_x^2+\sigma_y^2\right) - \frac{c}{2\sqrt{2}} \sigma_x^2 \sigma_y +\frac{\lambda_1}{2} \sigma_x^2\sigma_y^2 + \frac{1}{8} \left(2\lambda_1+\lambda_2\right)\sigma_x^4 \nonumber \\
&+& \frac{1}{4} \left(\lambda_1+\lambda_2\right) \sigma_y^4 - h_x \sigma_x - h_y \sigma_y. \label{eq:PotSU3}
\end{eqnarray}
The global minimization of the grand potential determines the $h_x$ and  $h_y$ as
\begin{eqnarray}
    h_x &=& m^2\sigma_x - \frac{{c}}{\sqrt{2}} \sigma_x \sigma_y - \lambda_1 \sigma_x \sigma_y^2 + \frac{1}{2}\left(2 \lambda_1 + \lambda_2\right) \sigma_x^3, \label{eq:hx}\\
h_y &=& m^2\sigma_x - \frac{{c}}{2\sqrt{2}} \sigma_x^2 -  \lambda_1\sigma_x^2 \sigma_y + \left( \lambda_1+ \lambda_2\right) \sigma_y^3. \label{eq:hy}
\end{eqnarray}
The tree-level masses of mesons are defined from the quadratic terms of the Lagrangian and are presented in detail in Appendix \ref{sec:AppB}.

The following section introduces the SU($4$) configuration.

\subsection{SU(4) configuration}
\label{sec:su4}

The tree-level mesonic potential for the scalar and pseudoscalar states in the SU(4) configuration can be derived from Eq. (\ref{eq:GenMesPot})
\begin{eqnarray}
U(\bar{\sigma})& =& 
\frac{m^2}{2}\bar{\sigma}_a^2 + \frac{1}{3} \left[ 
{\cal F}_{abcd}+{\cal G}_{abcd}
\right] \bar{\sigma}_a \bar{\sigma}_b \bar{\sigma}_c \bar{\sigma}_d 
- h_a \bar{\sigma}_a + \epsilon_a\sigma_a^2.
\label{eq:MesPotSU4}
\end{eqnarray}
The corresponding basis transformation to pure non-strange, strange, and charm fields, respectively, reads
\begin{eqnarray}\label{eq:eta_base}
  \left( 
\begin{array}{c}
 \sigma_x       \\
\sigma_y   \\
\sigma_c   
\end{array}
\right)   = 
\left( 
\begin{array}{ccc}
 \frac{1}{\sqrt{2}}    & \frac{1}{\sqrt{3}} & \frac{1}{\sqrt{6}}   \\
\frac{1}{2}    &  -\sqrt{\frac{2}{3}}           & \frac{1}{2\sqrt{3}}  \\
\frac{1}{2} & 0 & -\frac{\sqrt{3}}{2}         \\
\end{array}
\right) 
\left( 
\begin{array}{c}
 \sigma_0       \\
\sigma_8    \\
\sigma_{15}     
\end{array}
\right). 
\label{eq:SU4transition}
  \end{eqnarray}
The explicit chiral symmetry breaking parameters allow for the transformation of $h_0$, $h_8$, and $h_{15}$ using the same basis
\begin{eqnarray}
    H_{SU(4)} =T_0 h_0+ T_8 h_8 + T_{15} h_{15}\equiv \frac{1}{2}\left(
    \begin{array}{cccc }
 h_{x}     &0          & 0  &0\\
0    &  h_{x}           & 0 &0  \\
0 & 0 & \sqrt{2} h_{s}&0  \\
0 & 0 &0& \sqrt{2} h_{c}  
\end{array}\right).
\end{eqnarray}
Finally, the tree-level SU($4$) mesonic potential in new basis reads as
\begin{eqnarray}
 U(\sigma_x,\sigma_y, \sigma_c) &=&\frac{1}{2} m^2 \left(\sigma_x^2+\sigma_y^2+\sigma_c^2\right) -\frac{c}{4}\sigma_x^2\sigma_y\sigma_c \nonumber  +\frac{\lambda_1}{2} (\sigma_x^2\sigma_y^2+\sigma_x^2\sigma_c^2+\sigma_y^2\sigma_c^2)  + \frac{1}{8} \left(2\lambda_1+\lambda_2\right)\sigma_x^4 \nn \\
&+&  \frac{1}{4} \left(\lambda_1+\lambda_2\right) \sigma_y^4 +\frac{1}{4} \left(\lambda_1+\lambda_2\right) \sigma_c^4- h_x \sigma_x -  h_y \sigma_{y} - h_{c} \sigma_{c} +\epsilon_c \sigma_c^2.
\end{eqnarray}
The global minima, defined by the absence of partial derivatives related to $\sigma_x$, $\sigma_y$, and $\sigma_c$, define the values of $h_x, h_y, h_c$,
\begin{eqnarray}
h_x &=& m^2 \sigma_x - \frac{{c}}{2}  \sigma_x  \sigma_y  \sigma_c + \lambda_1  \sigma_x (\sigma_y^2 + \sigma_c^2 ) + \frac{1}{2}\left(2\lambda_1+\lambda_2\right)  \sigma_x^3, \\
h_y &=& m^2 \sigma_y - \frac{{c}}{4}  \sigma_x^2  \sigma_c + \lambda_1 \sigma_y(\sigma_x^2  +  \sigma_c^2) + \left(\lambda_1+\lambda_2\right)  \sigma_y^3, \\
h_c &=& m^2 \sigma_c + 2 \epsilon_c \sigma_c- \frac{{c}}{4}  \sigma_x^2  \sigma_y  + \lambda_1 \sigma_c ( \sigma_x^2  +\sigma_y^2)  + \left(\lambda_1+\lambda_2\right)  \sigma_{c}^{3}.
\end{eqnarray}
The tree-level masses of mesons, defined from the quadratic terms of the Lagrangian are presented in detail in Appendix \ref{sec:AppB}. As previously stated, the derivation for the masses of the charmed mesonic states requires to take into account the mass term $-2 \rm{Tr}[\epsilon \Phi^\dagger\Phi]$ \cite{Giacosa:2012cq}, where 
\begin{eqnarray}
    \epsilon =\left(
    \begin{array}{cccc }
0   &0          & 0  &0\\
0    & 0           & 0 &0  \\
0 & 0 & 0&0  \\
0 & 0 &0& \epsilon_c  
\end{array}\right), 
\end{eqnarray}
with $\epsilon_c = m_c^2$ and $m_c$ is the mass of the charm quark. This term provides an additional contribution specifically to the masses of charmed mesons. 

The formalism for the temperature dependence shall be elaborated in the following section. The derivation of analytical formulas for meson masses, with a focus on their behaviour at limited temperatures, will be presented.

\subsection{Finite Temperature Formalism}
\label{sec:finiteT}

To investigate how various in-medium modifications behave, we utilize the well-established Polyakov loop extension of the LSM. In this regard, we apply the chiral Lagrangian together with the Polyakov-loop potential, formulated as $ \mathcal{L} = \mathcal{L}_{chiral} - \mathbf{\mathcal{U}}(\phi, \phi^*, T)$. The grand potential can be derived then in the mean field approximation \cite{Tawfik:2019kaz}. Given the assumption of thermal equilibrium, the grand partition function is expressed via a path integral that includes the quark, antiquark, and meson fields
\begin{eqnarray}
\mathcal{Z} &=& \mathrm{Tr\, exp}[-\hat{\mathcal{H}}/T] = \int\prod_a \mathcal{D} \sigma_a \mathcal{D} \pi_a \int
\mathcal{D}q \mathcal{D} \bar{q} \mathrm{exp} \left[\int_x \mathcal{L}\right], 
\end{eqnarray}
where  $t$ is the time at which the system with volume $V$ evolves and $\int_x\equiv i \int^{1/T}_0 dt \int_V d^3x$.  The partition function can be derived using the mean field approximation \cite{Schaefer:2008hk,Schaefer:2006ds,Scavenius:2000qd}. In this context, the meson fields are substituted with their expectation values, specifically $\bar{\sigma_a}$, within the action \cite{Cheng:2007jq,Kapusta:2006pm}. Employing standard techniques \cite{Kapusta:2006pm}, the integration over the fermionic contributions can be performed. This leads to the derivation of the effective potential for the mesons
\begin{eqnarray}
\Omega(T)=\frac{-T \mathrm{ln}
\mathcal{Z}}{V}=U(\bar{\sigma})+\mathbf{\mathcal{U}}(\phi, \phi^*, T)+\Omega_{\bar{q}
q}, \label{potential}
\end{eqnarray}
where the fields $\phi$ or $\phi^*$ are a complex matrix of dimensions $N_f \times N_f$. This effective potential has three terms which can be elaborated as follows. 
\begin{itemize}
    \item The first term is the purely mesonic potential presented in previous sections that describes the mass spectra in vacuum. 
    \item The second term is the Poyakov loop potential, which in this study is used in classic logarithmic form \cite{Tawfik:2019rdd}. Additional information on the other potentials $\mathbf{\mathcal{U}}(\phi, \phi^*, T)$ can be found in Refs. \cite{Tawfik:2019tkp,Tawfik:2019kaz,Tawfik:2021eeb}.
    \item The third term is the mean-field quark-antiquark potential which gives an in-medium modification on masses
\begin{eqnarray}
        \Omega_{\bar{q}q} = - 2 \nu_c T\sum_f\int\frac{dp}{(2\pi)^3} E_f - 2 \nu_c T\sum_f\int\frac{dp}{(2\pi)^3}(\ln g^+_f +\ln g^-_f),
\end{eqnarray}
where $\nu_c=2\, N_c$ and $f$ run all flavours, $E_f = (p^2+m_f^2)^{1/2}$  is the flavor dependent single particle energy of quark with mass $m_f$  and  
\begin{eqnarray}
      g^{+}_f &=&   1+3(\phi+\phi^* e^{-E^+_{f}/T}) e^{-E^+_{f}/T}+e^{-3E^+_{f}}/T, \\
      g^{-}_f &=& 1+3(\phi^*+\phi e^{-E^-_{f}/T}) e^{-E^-_{f}/T}+e^{-3E^-_{f}}/T.
\end{eqnarray}
\end{itemize}
The quark and antiquark dispersion relations are ${\displaystyle E^{\pm}_{f}(T,\mu_f) = E_f \mp \mu_f}$, respectively. The quark masses are defined by values of the $\sigma$-fields:
\bea
    m_x = \frac{g}{2}\sigma_x, \qquad
    m_s = \frac{g}{\sqrt{2}}\sigma_y, \qquad
    m_c = \frac{g}{\sqrt{2}}\sigma_c. \nn
\eea
The masses of the various states can be determined from the second derivative of the grand potential with respect to the corresponding fields, evaluated at the potential minimum \cite{Tawfik:2014gga,Tawfik:2019rdd,Tawfik:2025hhr}. In the present work, these minima are defined by vanishing expectation values for all scalar, pseudoscalar, vector, and axial-vector fields \cite{Schaefer:2008hk}
\begin{equation}
(m^2_i)_{ ab} = \left.\frac{\partial^2 \Omega}{\partial\varphi_{i,a}  \partial \varphi_{i, b}}\right|_{\mathtt{min}}, \label{eq:GrndPtnSU3}
\end{equation}
where $\varphi_{i,a}$ and $\varphi_{i,b}$ are the corresponding mass fields of the $i$-th hadron state. 
Assuming that the contribution of the quark-antiquark potential to the Lagrangian vanishes in the vacuum, the mass matrix in hot and dense matter should be extended by the additional term \cite{Schaefer:2008hk}
\begin{equation}
(m_i^2)_{ab} = \left.\frac{\partial^2 U(\bar{\sigma}) (T, \, \mu) }{\partial \, \varphi_{i,a} \, \partial \, \varphi_{i,b}}\right|_{\rm min} + \left. \frac{\partial^2 \Omega_{q\bar{q}}(T, \, \mu) }{\partial \, \varphi_{i,a} \, \partial \, \varphi_{i,b} }\right|_{\rm min}, \label{eq:massmedium}
\end{equation}
where $i$ stands for (pseudo)scalar and (axial)vector mesons and $a$ and $b$ are integers ranging from $0$ to  ($N_f^2-1$). The first term in Eq. (\ref{eq:massmedium})  is related to the vacuum, where the meson masses are developed from the sigma fields. The second term gives the in-medium modification on the masses of various meson states 
\begin{eqnarray}
\frac{\partial^2 \Omega_{q\bar{q}}(T, \, \mu) }{\partial \, \varphi_{i,a} \, \partial \, \varphi_{i,b} } &=& \left. \nu_{c}\sum_{f}\int_0^{\infty} \frac{dp}{(2\pi)^3}  \right.  \frac{1}{2E_{ f}} \biggl[ (n^+_{f} + n^-_{{f}} ) \biggl( m^{2}_{f,a b} - \frac{m^{2}_{f,a}
 m^{2}_{f, b}}{2 E_{ f}^{2}} \biggr) 
\nn   \\ 
&+& (b^+_{f} + b^-_{{f}}) \biggl(\frac{m^{2}_{f,a}  m^{2}_{f, b}}{2 E_{f}\; T}
\biggr) \biggr],  \label{eq:ftmass}
\end{eqnarray} 

The expression Eq. (\ref{eq:ftmass}) contains some notations for the quark mass first derivative with respect to the meson fields   
$ m^2_{f,a} \equiv \partial m^2_f/\partial \varphi_{i,a}$ and the second derivative of the quark mass with respect to meson fields 
$ m^2_{f,{ab}} \equiv \partial m^2_f/\partial \varphi_{i,a} \partial  \varphi_{i,b}$. The values of these derivatives for SU(3) configuration can be found, for example, in Refs. \cite{Schaefer:2008hk,Gupta:2009fg} and for SU(4) case are presented in Appendix \ref{appndC}, Tab. \ref{tab:deriv}.
The notations $n^\pm_{f}$ and $b^\pm_{f}$ have the following definitions \cite{Tawfik:2014gga,Tawfik:2019rdd,Tawfik:2025hhr}
\begin{eqnarray}
n^+_{f} &=& \frac{\phi e^{-\,E^+_{f}/T} + 2 \phi^* e^{-2\,E^+_{f}/T} + e^{-3\,E^+_{f}/T}}{1+3(\phi+\phi^* e^{-E^+_{f}/T}) e^{-E^+_{f}/T}+e^{-3E^+_{f}/T}}, \\
n^-_{{f}} &=& \frac{\phi^* e^{-E^-_{{f}}/T} + 2 \phi e^{-2E^-_{{f}}/T} + e^{-3E^-_{{f}  }/T}}{ 1+3(\phi^*+\phi e^{-E^-_{{f}}/T}) e^{-E^-_{{f}}/T}+e^{-3E^-_{{f}}/T}}.
\end{eqnarray}
The normalization factors for quarks and antiquarks are $b^{\pm}_{f}=3 (n^{\pm}_{f})^2 - c^{\pm}_{f}$ where \cite{Tawfik:2014gga,Tawfik:2019rdd,Tawfik:2025hhr} 
\begin{eqnarray} 
c^+_{f} &=& \frac{\phi e^{-\,E^+_{f}/T} +4 \phi^* e^{-2\,E^+_{f}/T} +3 e^{-3\,E^+_{f}/T}}{1+3(\phi+\phi^* e^{-E^+_{f}/T})\, e^{-E^+_{f}/T}+e^{-3E^+_{f}/T}}, \\
c^-_{{f}} &=&  \frac{\phi^* e^{-E^-_{{f}}/T} + 4 \phi e^{-2E^-_{{f}}/T} +3 e^{-3E^-_{{f}}/T}}{ 1+3(\phi^*+\phi e^{-E^-_{{f}}/T})\, e^{-E^-_{{f}}/T}+e^{-3E^-_{{f}}/T}}.
\end{eqnarray}

The numerical results and the procedure for parameter fitting are introduced in the following section. 

\section{Results}
\label{sec:rslts}

The Lagrangian of eLSM, given in Eqs. (\ref{eq:Lagrangian}-\ref{eq:LagrSum}), contains a large number of parameters: 
\bea
m_0^2, m_1^2, c, \delta_x, \delta_y, \delta_c,  g_1,  g_2,  g_3, g_4, g_5, g_6,  h_{x},  h_{y}, h_{c}, h_1, h_2, h_3, \lambda_1, \lambda_2. \nn
\eea
In this study, the coupling of the glueball to other mesons is neglected. Furthermore, the parameters  $g_2, g_3, g_4, g_5$ and $g_6$  in Eq. (\ref{eq:vector_nonets}) are not discussed in this work and therefore are not considered in the fit.  Fitting these parameters requires fixing some meson masses and decay constants. The remaining masses and decay constants are then obtained from the fit itself. 

The condensate values in vacuum are computed from the meson decay constants using the partially conserved axial-vector current relation (PCAC) \cite{Lenaghan:2000ey}
\begin{equation}
f_a = d_{aab}{\sigma_b},
\end{equation}
with summation over $b$ and  $d_{abc}$ is the standard symmetric structure constants of SU($N$). The $\sigma_x$ and $\sigma_y$ are defined by the pion and kaon decay constants. The value of $\sigma_c$ can be expressed in terms of the D-meson decay constant or, alternatively,  $\eta_c$ meson decay constant as $f_{\eta_c} = 2 \sigma_c$  \cite{Tawfik:2025hhr,Diab:2025blc,Eshraim:2014eka}
\bea
f_\pi = \frac{\sigma_x}{Z_\pi}, \qquad
f_K = \frac{\sqrt{2} \sigma_y +\sigma_x}{2 Z_K},\qquad
f_D = \frac{\sqrt{2}\sigma_c + \sigma_x}{2 Z_D}. \nn
\eea
As justified in Appendix \ref{sec:AppB}, the factors $Z_i$ are included in the expressions above. Table \ref{tab:expdata} presents the experimental data used in the fit alongside the corresponding best-fit results. We conclude that the estimation grounded in SU($4$) eLSM is more consistent with experimental values than that grounded in SU($3$) eLSM. This finding draws a key conclusion that an increase in quark degrees of freedom enhances the simulation accuracy of meson masses.

\begin{table}[h]
    \centering
    \begin{tabular}{|c|c|c|c|}
    \hline
         Observable &  Experiment (MeV) & SU(3) Fit (MeV)& SU(4) Fit (MeV)\\
         \hline
    $f_\pi$ & 92.2$\pm$ 4.6& 95.02 &92.0\\
    $f_K$ & 110.4$\pm$ 5.5& 106.9 &109.1\\
    $f_D$ & 203$\pm$ 5.5& -  &225.0\\
    $m_\pi$ & 137.3$\pm$ 6.9& 139.85&140.79  \\
    $m_K$ & 495.6$\pm$ 24.8& 420.49 &490.3\\
    $m_\eta$ & 547.9$\pm$ 27.4& 140.0 (531.9.0)& 140.79(531.13) \\
    $m_\eta'$ & 957.8$\pm$ 47.9& 644.93 (965.44)&640.75(965.56)\\
    $m_\phi$ & 1019.5$\pm$ 51& 1019.19 & 1014.9 \\
    $m_\rho$ & 775.5$\pm$ 38.8& 770.11 &749.62\\
    $m_{f_1}$ & 1426.4$\pm$ 71.3& 1424.77 &1460.39\\
    $m_{a_1}$ &1230$\pm$ 62&1069.76 &1078.68\\
    $m_{k_1}$ &1253$\pm$ 7&1253.46 &1276.45\\
    $m_{\sigma}$ &500-1200&600 (700) &745(700)\\
    $m_{D_1}$ &2420$\pm$ 9&- &2615.77\\
    \hline
\end{tabular}
    \caption{The data used for parameters fitting. In brackets are shown results for fit with $c\neq 0$ (if differ). }
    \label{tab:expdata}
\end{table}
 
The rest of the parameters are defined as follows:
\begin{itemize}
\item $\delta_x, \delta_y, \delta_c$ define explicit symmetry breaking in the vector and axial-vector channels, which arises from non-vanishing quark masses. In this regard, we find the following
\bea
    \Delta_{SU(3)} &=& T_0 \delta_0+ T_8 \delta_8\equiv \frac{1}{2}\left(
    \begin{array}{ccc}
 \delta_{x}     &0          & 0  \\
0    &  \delta_{x}           & 0   \\
0 & 0 & \sqrt{2} \delta_{s}  
\end{array}\right), \\
    \Delta_{SU(4)} &=& T_0 \delta_0+ T_8 \delta_8 + T_{15} \delta_{15}\equiv \frac{1}{2}\left(
    \begin{array}{cccc }
 \delta_{x}     &0    & 0  &0\\
0    &  \delta_{x}   & 0 &0  \\
0 & 0 & \sqrt{2} \delta_{s} & 0  \\
0 & 0 &0& \sqrt{2} \delta_{c}
\end{array}\right).
\end{eqnarray}
The transition from the original octet-singlet basis to the non-strange, strange, and charm quark flavor basis is performed using the transformation matrices in Eqs. (\ref{eq:SU3transition}) and (\ref{eq:SU4transition}). These symmetry-breaking parameters are proportional to the squares of the corresponding quark masses: $\delta_x \propto  m_x^2$,  $\delta_y \propto m_y^2$, and $\delta_c \propto m_c^2$. In the isospin-symmetric limit, we set $\delta_x=0$ and determine the remaining parameters $\delta_y$ and $\delta_c$ from the fit.  The quark masses are defined by the Yukawa coupling $g$ and the corresponding condensates. The Yukawa coupling $g$ is fixed from the non-strange constituent quark mass as $g = 2 m_q/{\sigma}_x$ where $m_q = 0.3$ GeV (see Table \ref{tab:params}).

\item The parameters $h_{x}$,  $h_{y}$, $h_{c}$ define the  Explicit Chiral Symmetry Breaking (ESB) in the (pseudo)scalar sector via the term $\rm{Tr}[H(\Phi +\Phi^\dagger)]$ and they are defined by the global minimum of the mesonic potential of the model (see Section \ref{sec:Lagrangian}).
\item The parameters $m_1^2, g_1$, $h_1, h_2, h_3$ define the vector and axial-vector sectors. As can be seen from the mass expressions in Tabs. \ref{tab:su3masses} and \ref{tab:su4noncahrmed}, the meson masses depend more significantly on the value of $c_1$, which can be written as
\begin{eqnarray}
        c_1 &=& m_1^2 + \frac{h_1}{2}(\sigma_x^2+\sigma_y^2) \qquad\qquad\qquad\, SU(3),\\
        c_1 &=& m_1^2 + \frac{h_1}{2}(\sigma_x^2+\sigma_y^2+\sigma_c^2)\qquad\qquad SU(4),
\end{eqnarray}
so that we can reduce the fit to  $g_1, h_2, h_3, c_1$.  The parameters obtained from the best fit are listed in Tab. \ref{tab:params}.

\item The parameters $m_0^2, c = 0, \lambda_1, \lambda_2$ define the scalar-pseudoscalar sector. The parameter $c$ define the presence of the $U(1)_A$ anomaly. For the case when $U(1)_A$ anomaly is absent at $c=0$. Accordingly, the mass of $\eta'$ meson is identical to the pion mass. The parameter $\lambda_2$ is given by the kaon and pion masses. The well-known classical relation for $\lambda_2$ reads
\begin{eqnarray}
\lambda_2 = \frac{M_K^2-M_\pi^2}{(2 f_K-f_\pi)(f_K-f_\pi)}.
\end{eqnarray}
It is identical for both the SU(3) and SU(4) configurations. The parameters $\lambda_1$ and $m_0$  then can be obtained  using  equations for masses of the scalar sigma-meson $m_{\sigma}$ and the pion mass $m_\pi$.

\item In the presence of $U(1)_A$ anomaly, i.e., $c\neq 0$, determining the model parameters requires a system of equations constrained by the following experimental inputs: \begin{itemize}
\item the pion and kaon masses, 
\item the averaged squared mass of $\eta$ and  $\eta'$ mesons ($m^2_\eta + m^2_{\eta'} = m^2_{\eta_N}+ m^2_{\eta_S}$), and 
\item the mass of the scalar sigma-meson. \end{itemize}
Using this solution, we can then predict the parameters, the masses of the scalar mesons  $m_{a_0}$ , $m_{f_0}$, as well as  the scalar and pseudoscalar mixing angles $\theta_S$, $\theta_P$.
\end{itemize}

The parameters $g_1$, $c_1$, $h_2$, $h_3$, $c$, $m_0^2$, $\lambda_1$, $\lambda_2$ and $g$ for both SU(3) and SU(4) configurations are specified in Tab. \ref{tab:params}. It is important to point out that the vacuum masses for the other mesons, which aren't included in the fit, can be found in Table \ref{tab:mesonMasses} in Appendix \ref{sec:AppB}.

\begin{table}[htb]
    \centering
    \begin{tabular}{|c|c|c|c|c|c|c|c|c|c|}
    \hline
    & $g_1$& $c_1$ (GeV$^2$)& $h_2$ & $h_3$ & c  & $m_0^2$ (GeV$^2$)& $\lambda_1$& $\lambda_2$& g  \\
    \hline
          \multirow{2}{*}{SU(3)}  & \multirow{2}{*}{5.58} & \multirow{2}{*}{0.44} & \multirow{2}{*}{40.975} & \multirow{2}{*}{-15.216} & 0 & -$(0.282)^2$& -10.515 & 55.576&\multirow{2}{*}{5.5} \\
            &  &  &  &  & 2.828 & -$(0.258)^2$ & 3.15 & 36.253& \\
    \hline
           \multirow{2}{*}{SU(4)}  & \multirow{2}{*}{5.632} & \multirow{2}{*}{0.395} & \multirow{2}{*}{39.215} & \multirow{2}{*}{-13.907} & 0 & -(0.135)$^2$& -2.115 & 45.54 &\multirow{2}{*}{5.217} \\
              &  &  &  &  & 11.24 & -(0.707)$^2$& 4.055& 29.19 & \\
    \hline
    \end{tabular}
    \caption{The best fit of the parameters $g_1$, $c_1$, $h_2$, $h_3$, $c$, $m_0^2$, $\lambda_1$, $\lambda_2$ and $g$.}
    \label{tab:params}
\end{table}

\begin{figure}[htbp]
\centerline{
\includegraphics[width = 0.56\textwidth]{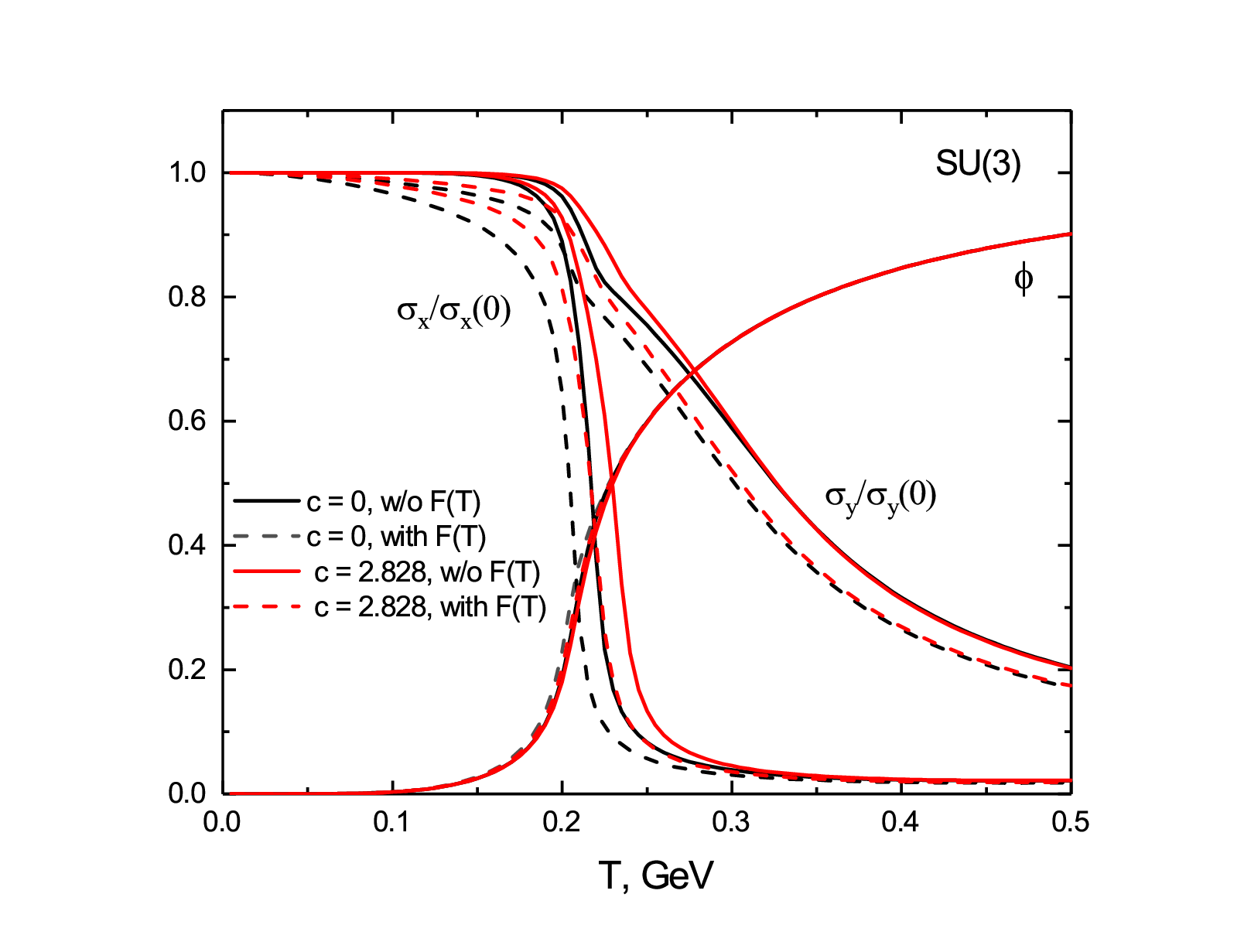}
\includegraphics[width = 0.56\textwidth]{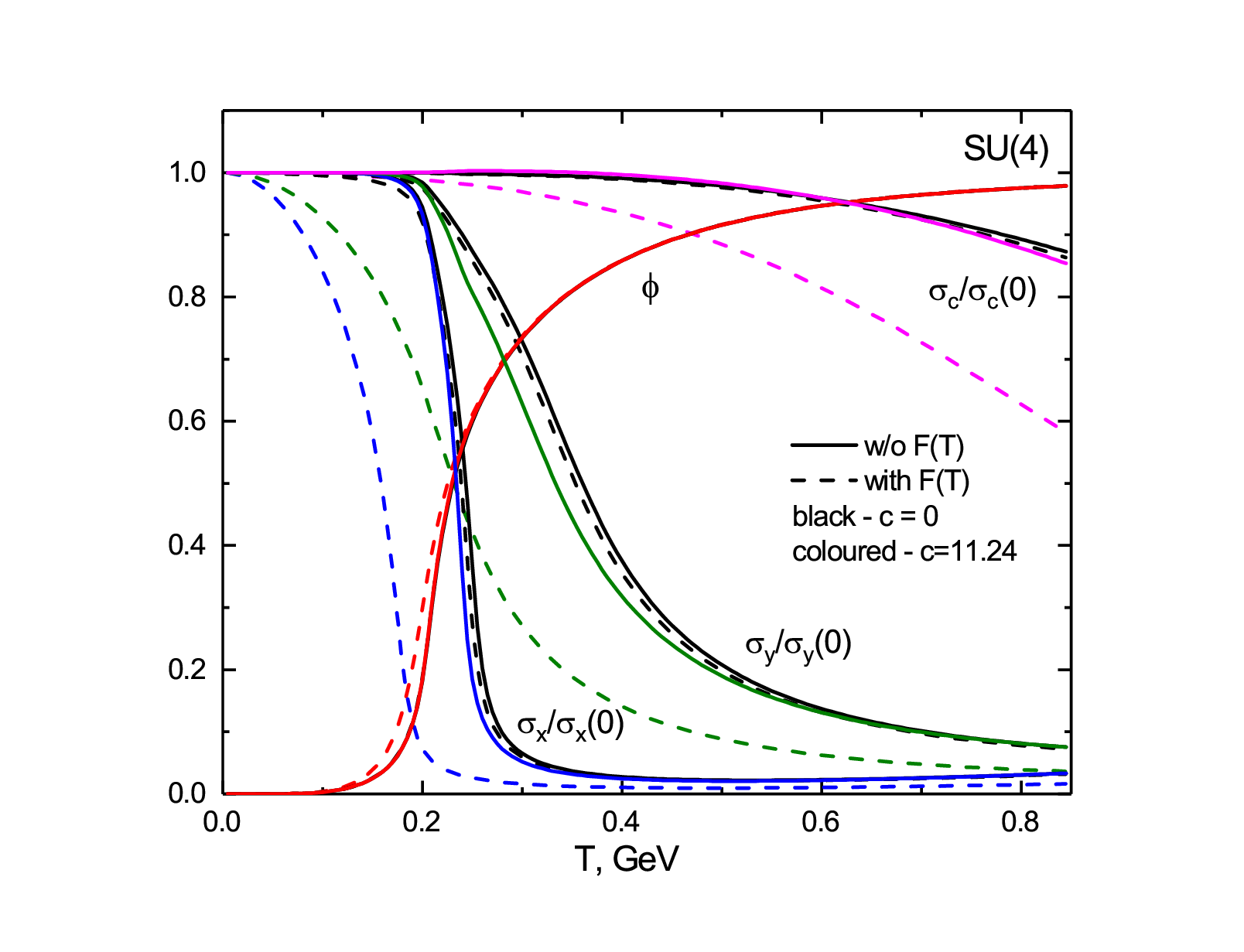} }
\caption{The quark condensates for SU(3) (shown in left panel) and SU(4) (presented in right panel) are displayed as a function of temperature in GeV units. The solid curves indicate calculations performed at $c=0$ with a vanishing $F(T)$, while the dashed curves represent calculations at $c=0$ with a finite $F(T)$. The dash-dotted curves illustrate the calculations at finite $c$ and vanishing $F(T)$. Calculations conducted at finite $c$ and finite $F(T)$ are depicted with dotted curves. All calculations are executed at $T_0=T_c=0.27$ GeV.}
\label{fig:condens}
\end{figure}

The Polyakov potential, as introduced in Section \ref{sec:finiteT}, depends on the parameter $T_0$, which represents the deconfinement critical temperature in the pure gauge sector with $T_0=0.27$ GeV.  Following the argument from the introduction and our previous research \cite{Tawfik:2025hhr}, we incorporate an additional temperature dependence,  making the substitution of the  parameter $m^2_0$ by  $\tilde{m}^2_0$ with 
\begin{equation}
    \tilde{m}^2_0 = m^2_0 \left(1- \frac{T^2}{T_c^2}\right), \label{eq:factor}
\end{equation} 
where $T_c$ is proportional to the  fundamental quantities of QCD, the scale for the hadron-quark confinement, i.e., $\Lambda_{\rm{QCD}}$, we keep $T_c$ for the first step as $0.27$ GeV. 
Figure \ref{fig:condens} shows the temperature dependence of quark condensates in the eLSM configuration with and without the factor $F(T) = \left(1- T^2/T_c^2\right)$ in Eq. (\ref{eq:factor}), and $T_0=T_c=0.27$ GeV. The Figure clearly shows the restoration of chiral symmetry in the light quark sector ($\sigma_x$) and  strange ($\sigma_y$) sector at higher temperature, but and charm quark condensates ($\sigma_c$) stays modestly temperature dependent over a large temperature range.

The temperature of the deconfinement transition  $T_d$ can be defined as the inflection point $\rm{max}\left|\partial \Phi/\partial T\right|_{\mu=const}$  and the pseudo-critical temperature of the chiral phase transition $T_\chi$ can be defined as ${max}
\left|\partial \sigma_f/\partial T\right|_{\mu=const}$. Table \ref{tab:Tc} presents the obtained values of $T_\chi$ in the light sector for all considered modifications of the model in the case $T_0=T_c=0.27$ GeV. The temperature behavior of the condensates is mainly governed by the background gauge field $\Phi$ and the value of the parameter $T_0$ plays crucial role in determining the pseudo-critical temperature of the chiral phase transition, $T_\chi$. It is clearly seen, that the resulting pseudo-critical temperature is higher than that suggested by the latest Lattice QCD results \cite{Gavai:2024mcj}. One way to improve this situation is to rescale the value of $T_0$, thereby simulating the back-reaction of quarks \cite{Schaefer:2007pw}. Nevertheless, it is seen clearly, that the factor $F(T)$, Eq. (\ref{eq:factor}), effectively plays the role of such a back-reaction, leading to a reduction of $T_\chi$ for both SU(3) and SU(4) configurations.

\begin{table}[t]
    \centering
    \begin{tabular}{|c|c|c|c|c|}
    \hline
         & $c = 0$ & $c = 0$ with F(T)& $c\neq 0$ & $c\neq 0$, with F(T)  \\
         \hline
        SU(3) & 0.2175 & 2.075 & 0.2325 & 0.2175 \\
        SU(4) & 0.2475 & 2.425 & 0.2375 & 0.1725 \\
        \hline
     \end{tabular}
    \caption{The pseudo-critical temperature of the chiral phase transition ($T_\chi$) for SU(3) and SU(4) configurations with and without the factor $F(T) $ with $T_0=T_c=0.27$ GeV.}
    \label{tab:Tc}
\end{table}

\begin{figure}[htbp]
\centerline{
\includegraphics[width=0.56\textwidth]{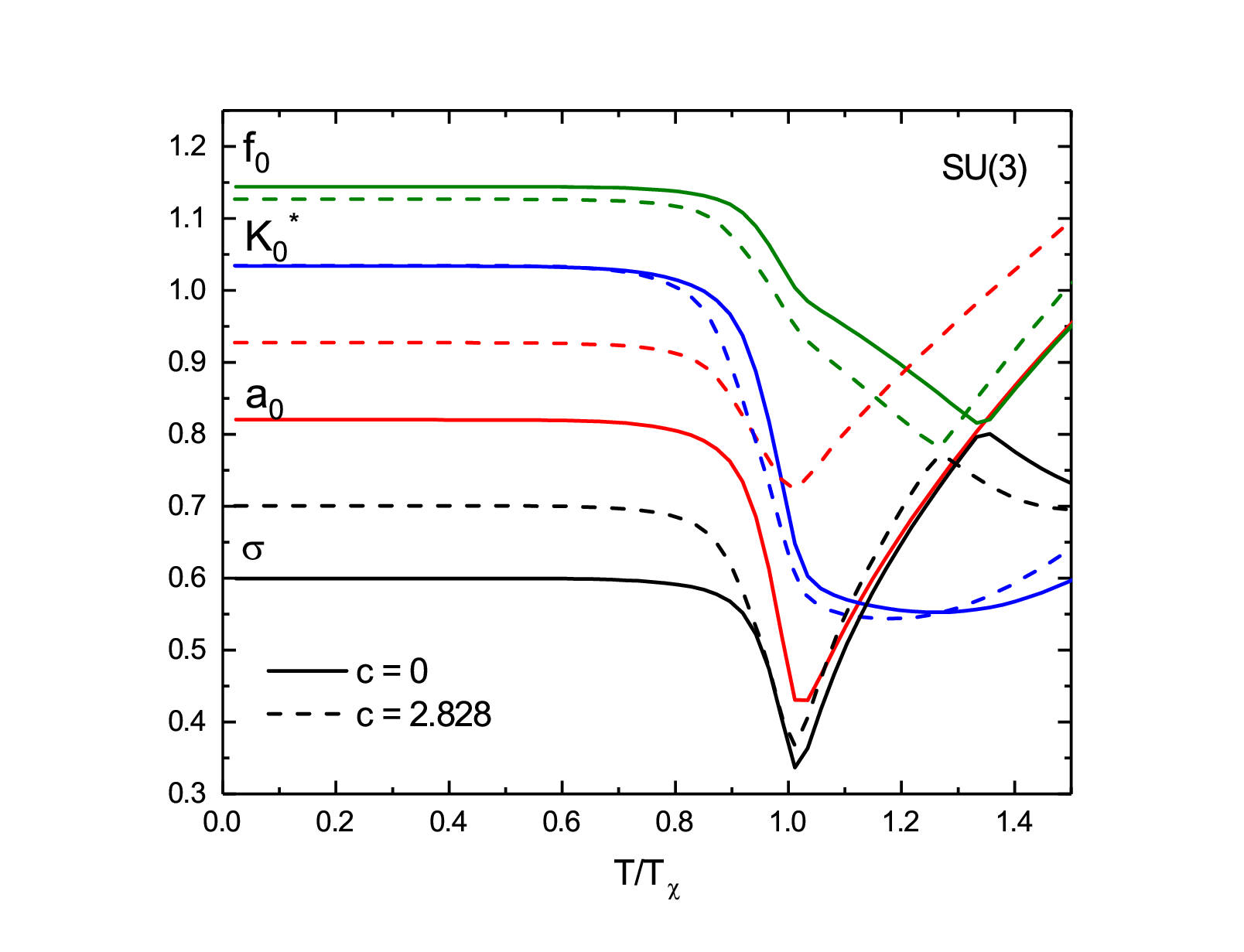}
\includegraphics[width=0.56\textwidth]{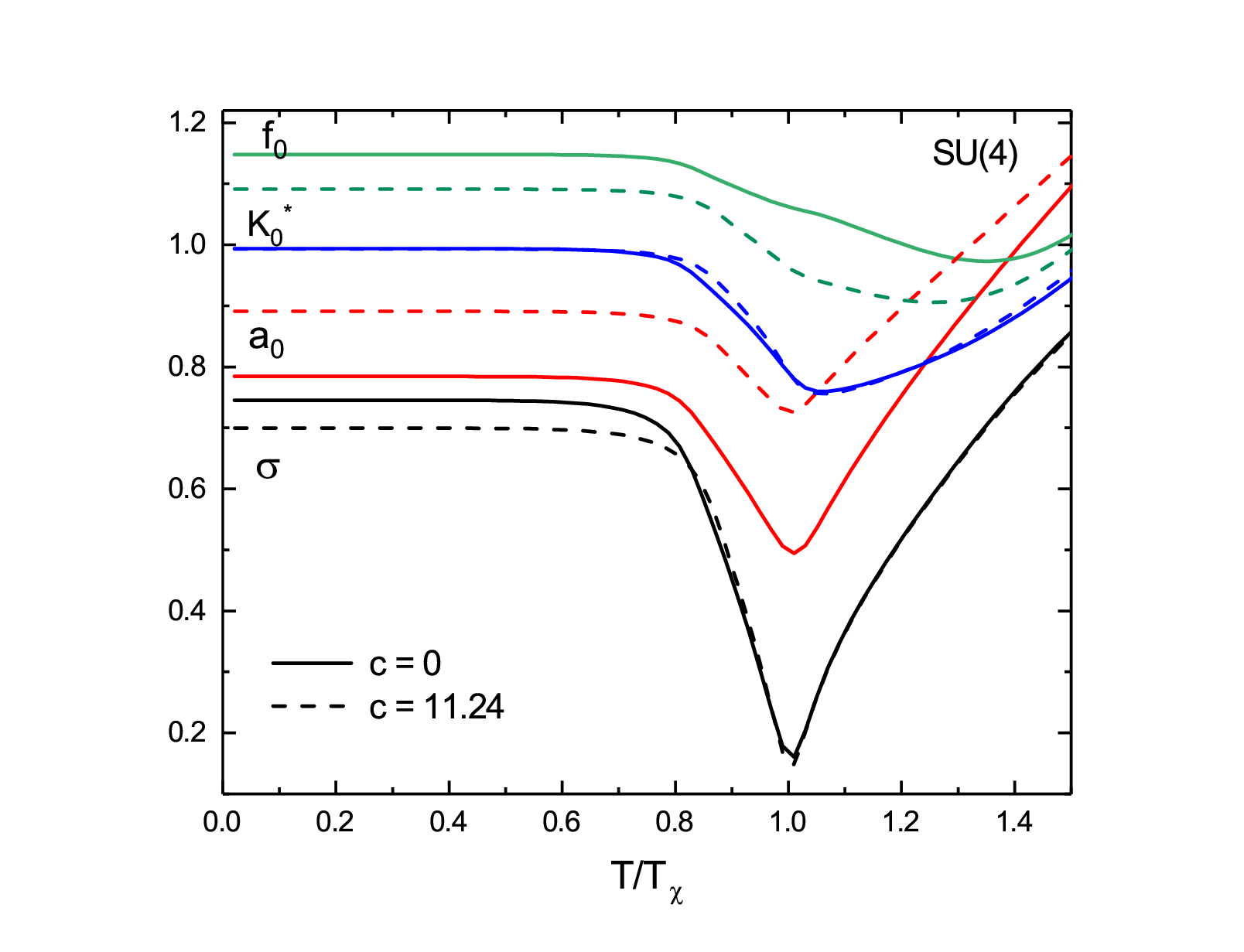} }
\caption{Scalar mesons for SU(3) (left panel) and SU(4) (right panel). Meson masses are plotted both for  $c=0$ (solid lines),  $c \neq 0$ (dashed lines) cases  without involving the factor $F(T) = (1-T^2/T_0^2)$. }
\label{fig:scalars}
\end{figure}

\begin{figure}[htbp]
\centerline{
\includegraphics[width = 0.56\textwidth]{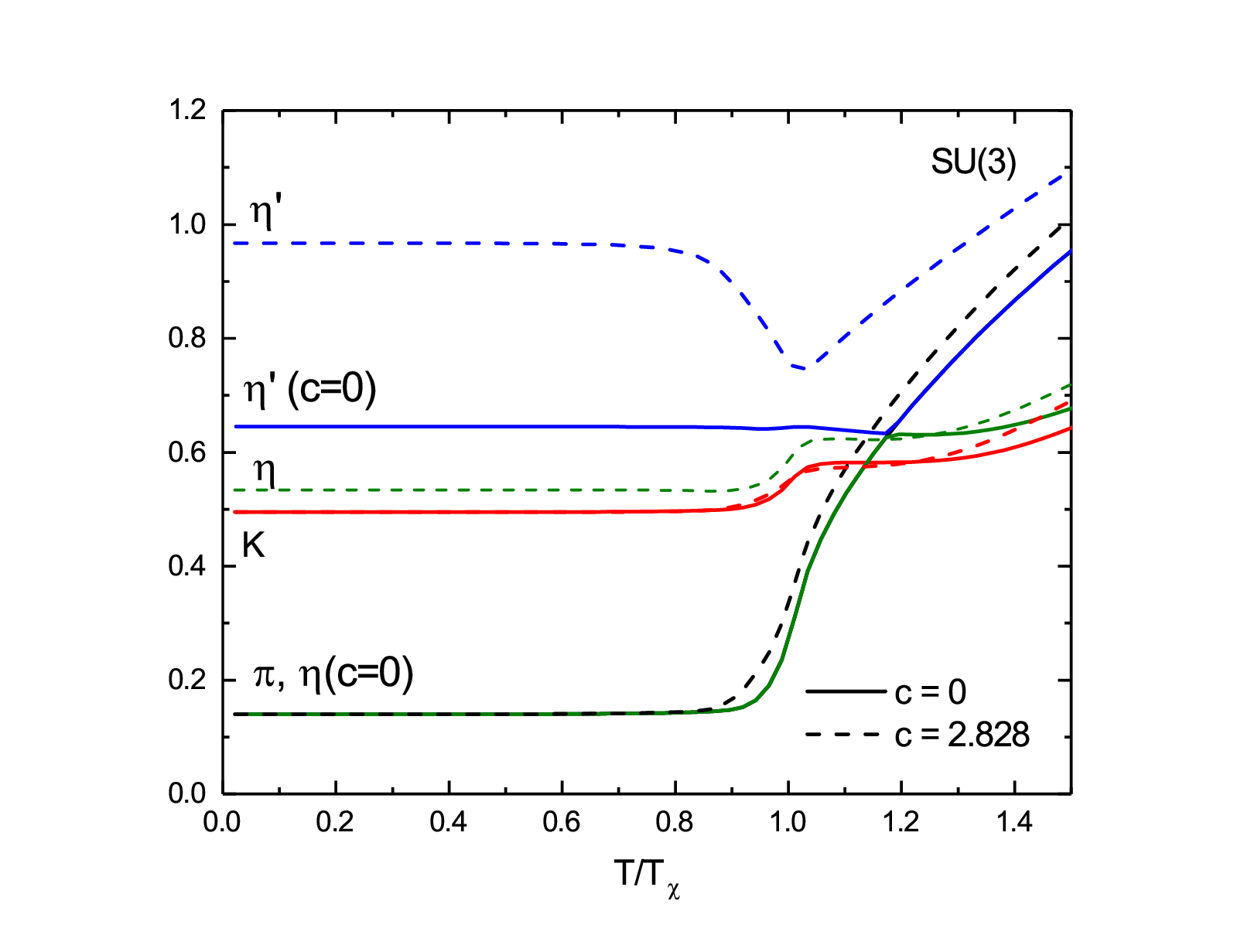}
\includegraphics[width = 0.56\textwidth]{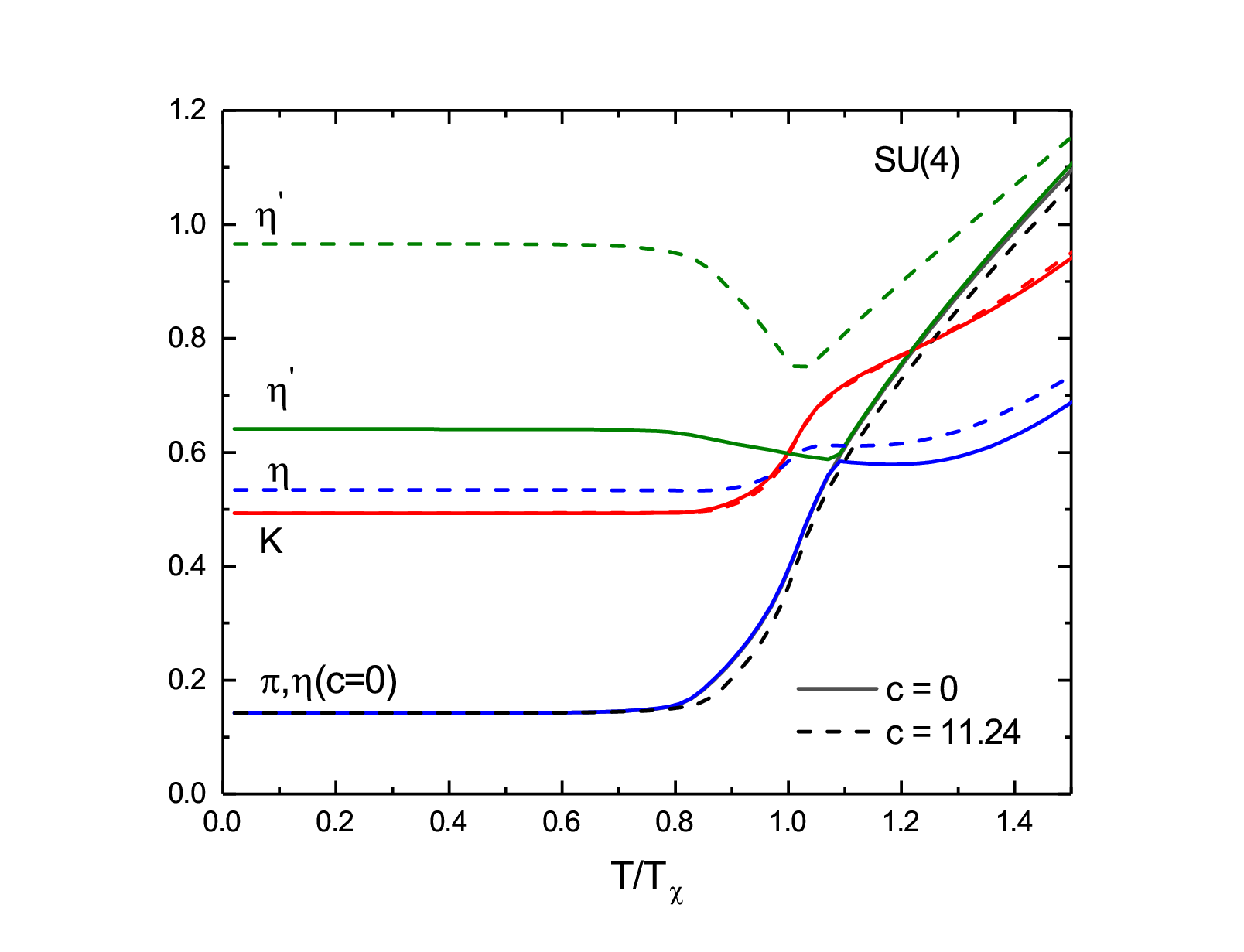} }
\caption{Masses of pseudo-scalar mesons for SU(3) (left panel) and SU(4) (right panel) cases. Meson masses are plotted for  $c=0$ (solid lines) and  $c \neq 0$ (dashed line) cases. }
\label{fig:pseudoscalars}
\end{figure}

\begin{figure}[htbp]
\centerline{
\includegraphics[width=0.56\textwidth]{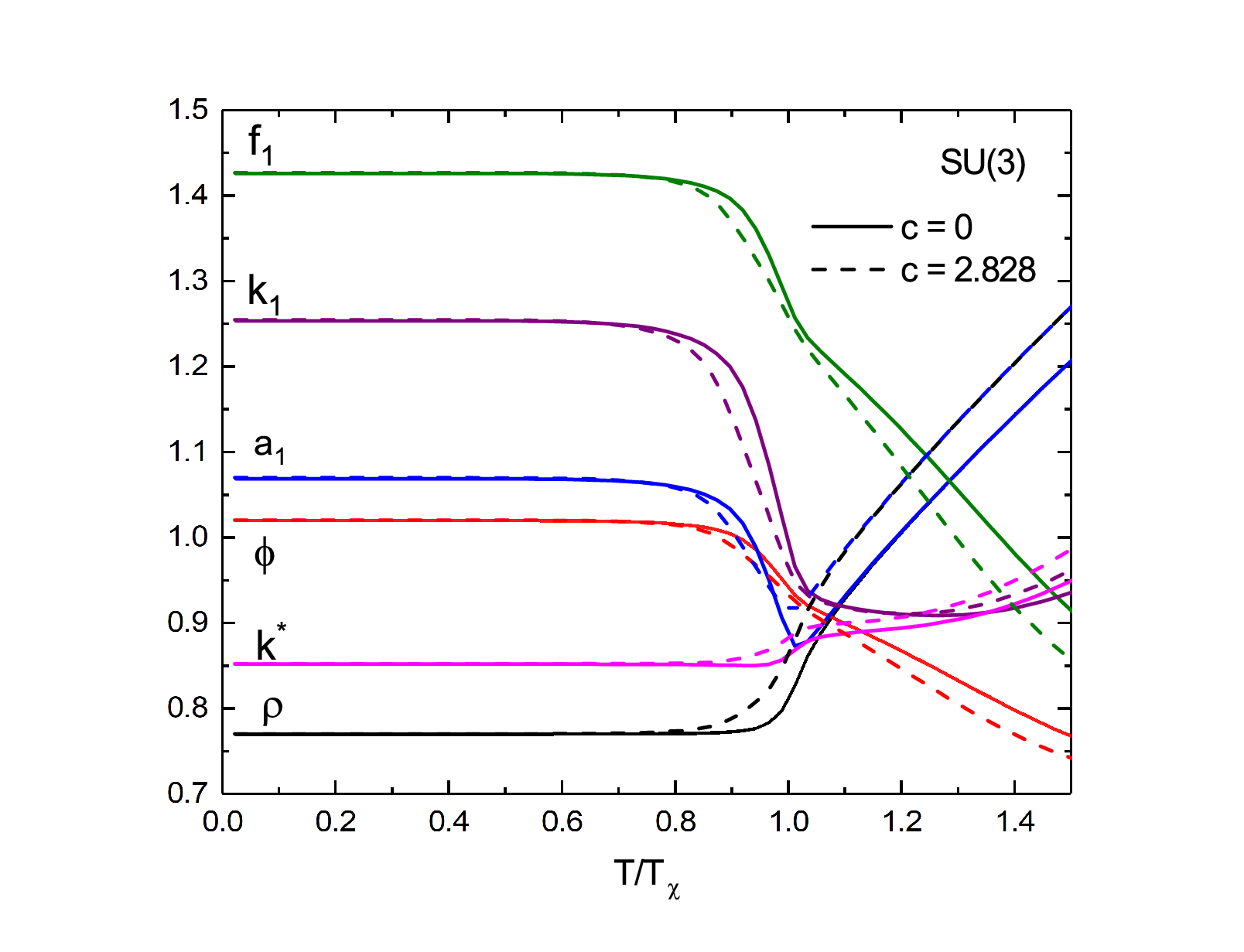}
\includegraphics[width=0.56\textwidth]{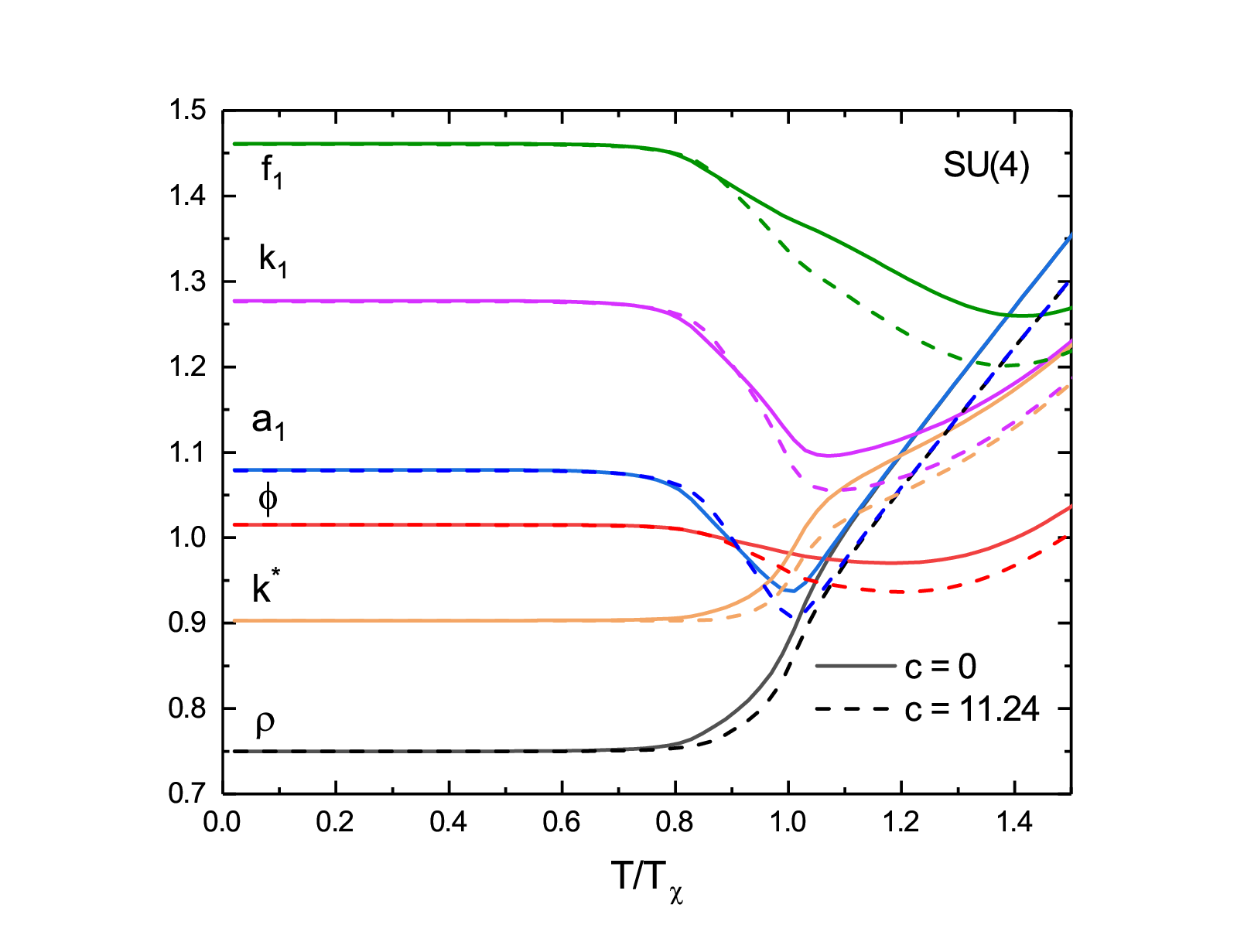} }
\caption{Vector and axial-vector mesons for SU(3) (left panel) and SU(4) (right panel). In the left panel, the different meson masses are illustrated for both the cases of $c=0$ (solid lines) and $c\neq 0$ (dashed lines), whereas in the right panel, the results are obtained at a finite value of $c$. }
\label{fig:vectors}
\end{figure}

\begin{figure}[htbp]
\centerline{
\includegraphics[width=0.56\textwidth]{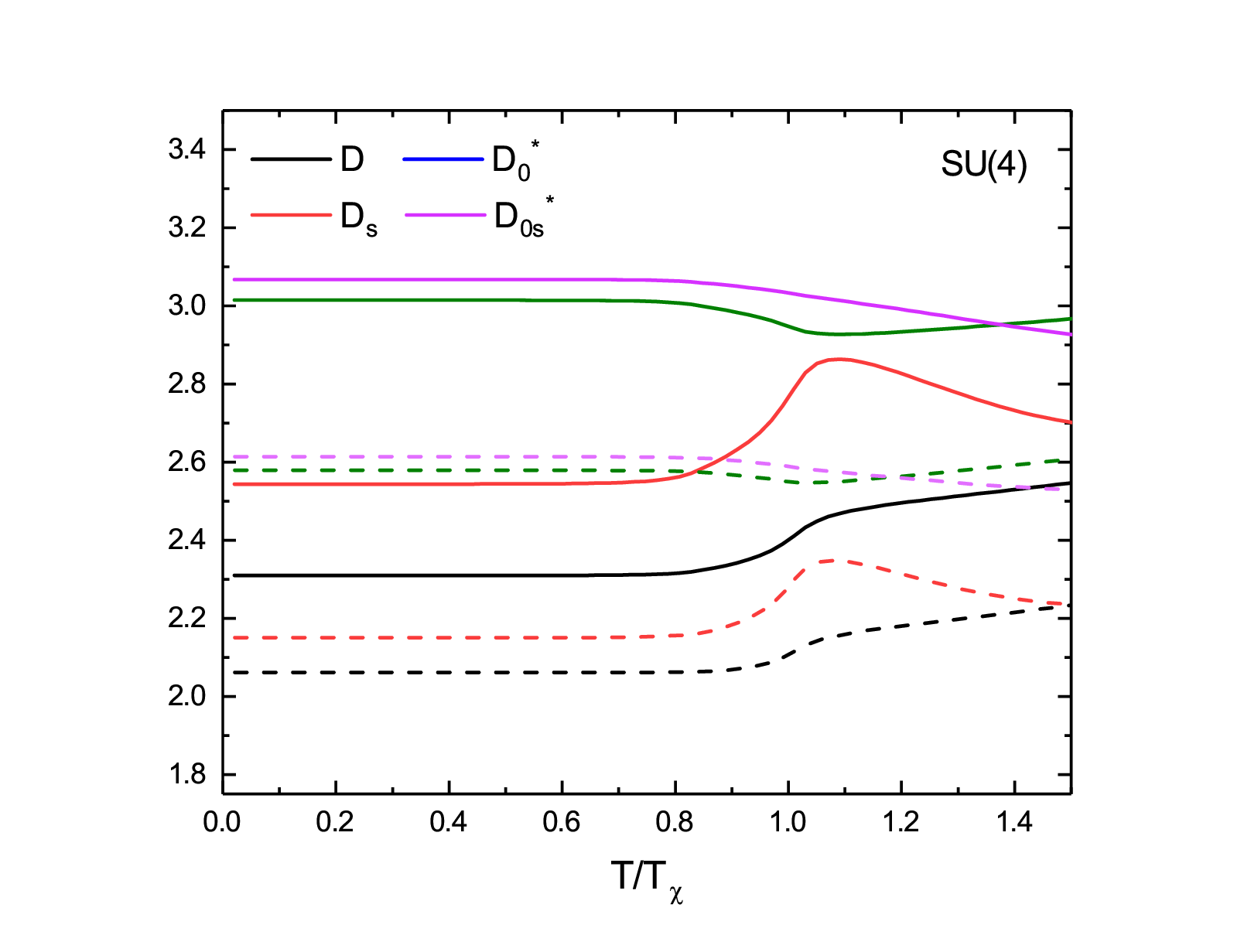}
\includegraphics[width=0.56\textwidth]{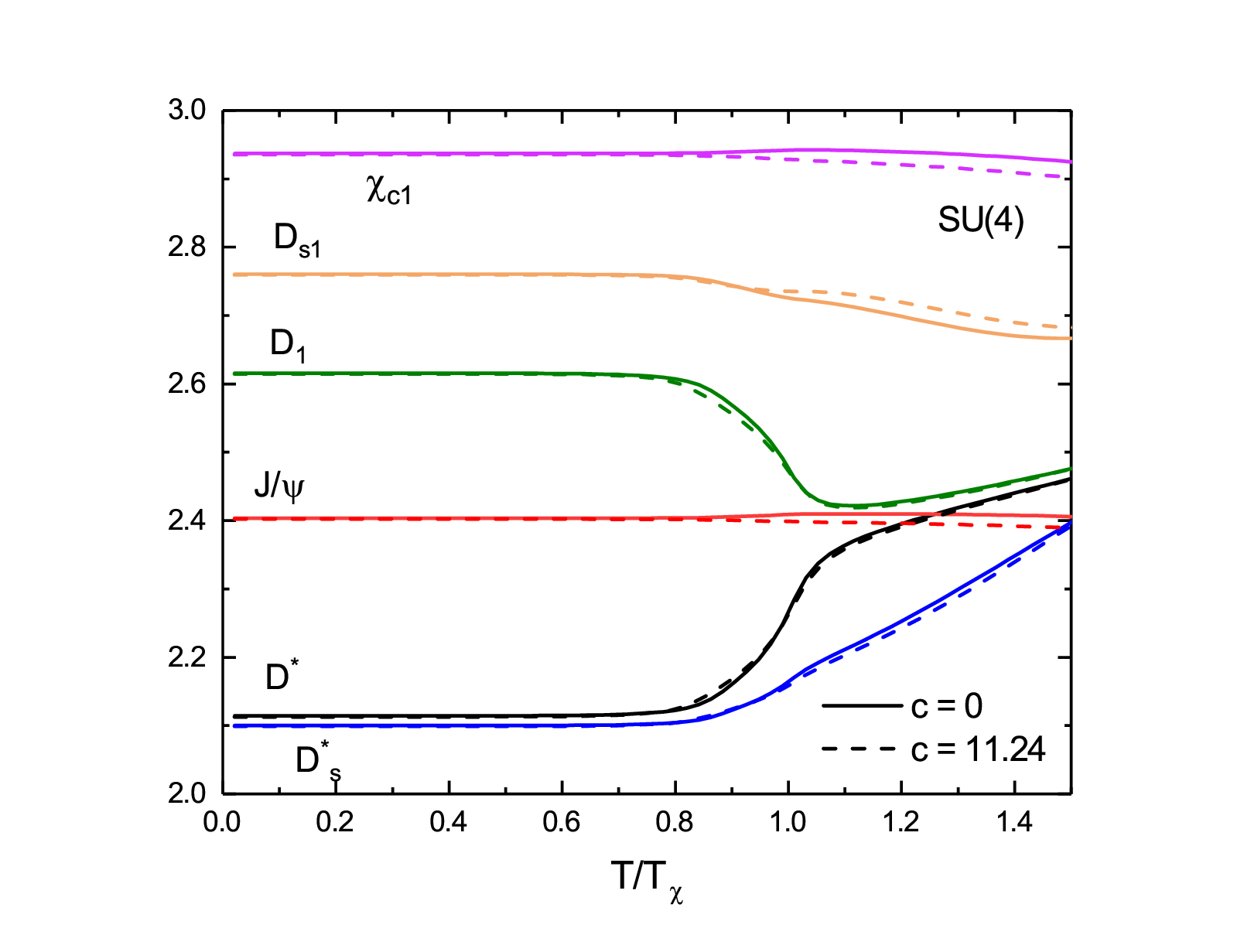} }
\caption{Left panel: scalar and pseudo-scalar mesons with open charm and quarkonia. Right panel: vector and axial-vector mesons with open charm and quarkonia. Meson masses are plotted both for  $c=0$ (solid lines),  $c \neq 0$ (dashed lines) cases. }
\label{fig:charmed}
\end{figure}

The illustrations of the thermal evolution of the meson mass spectrum are presented in Fig. \ref{fig:pseudoscalars} for the pseudoscalar mesons, in Fig. \ref{fig:scalars} for scalar mesons, in Fig. \ref{fig:vectors} vector and axial-vector mesons and in Fig. \ref{fig:charmed} for open/hidden charmed mesons. All are given at vanishing chemical potential $\mu_f$ without the factor F(T). For the pseudoscalar and scalar charmed mesons we presented only $D$, $D_s$, $D^{*}_{0}$, $D^{*}_{s0}$, states, because  masses of  $\eta_c$ meson with $m_{\eta_c} = 4.844$ GeV and  $\chi_{c0}$-meson with $m_{\chi_{c0}} = 3.551$ GeV behave as a constant as functions of temperature in the considered temperature range. These figures allow us to conclude that while various mesonic states display specific patterns in their mass evolution with temperature, their dissociation tends to happen within a similar temperature range close to the critical temperature, with minor variations depending on the type of meson. On the other hand, the quarkonium states seem to be largely unaffected by temperature changes.

\section{Conclusion}
\label{sec:cnls}

Using the Polyakov-loop extended Linear-Sigma Model, we studied the temperature behavior of mesonic states and of the characteristic fields $\sigma_x, \sigma_y, \sigma_c  $, which are the order parameters of the chiral phase transition.  Our results are consistent to the standard picture and demonstrate that  temperature significantly influences the light quark condensates $\sigma_x$, while the strange quark sector  $\sigma_y$ undergoes the chiral symmetry restoration at higher temperature. In contrast, the charm quark condensate $\sigma_c$ is almost independent of temperature. This observation suggests that the pseudo-critical temperature of the chiral phase transition increases when moving from light to strange to charm quark degrees of freedom. 
	
As can be seen from Table~\ref{tab:Tc}, the pseudo-critical temperature of the chiral phase transition in the light sector is higher than that predicted by Lattice QCD, $T_\chi \in [0.150, 0.170]$ MeV \cite{Gavai:2024mcj}. This feature is common in quark–meson models that incorporate the Polyakov loop (see, e.g., Ref.~\cite{Friesen:2016uof} for the PNJL model, Ref.~\cite{Schaefer:2007pw} for the quark–meson model, and Ref.~\cite{Kovacs:2016juc} for the eLSM). One possible way to improve this situation is to rescale the parameter $T_0$, thereby simulating the back-reaction of quarks \cite{Schaefer:2007pw}. In the present work we also propose to account for this back-reaction by introducing a temperature dependence of the parameter $m_0^2$, which affects the condensates. As shown in Table~\ref{tab:Tc}, this modification reduces the value of the pseudo-critical temperature, with the effect of the factor $F(T)$ being more pronounced in the SU(4) configuration. Nevertheless, it is possible to choose the parameters $T_0$ and $T_c$ in such a way as to reproduce the Lattice QCD results. The corresponding parameter sets are presented in Table~\ref{tab:Tc2} below.

\begin{table}[t]
	\centering
	\begin{tabular}{|c|c|c|c|c|c|}
		\hline
		& GeV & $c = 0$ &  $c\neq 0$ &$c = 0$ with F(T)&$c\neq 0$, with F(T)  \\
		\hline
		\multirow{3}{*}{SU(3)} & & \multicolumn{4}{c|}{$T_0 = 0.178, T_c = 0.156$}\\ \cline{3-6}
	    & $T_\chi$ &0.198 &0.208&0.153 &0.165 \\ 
		& $T_d$ &0.143 &0.148&0.143&0.143 \\ \hline
		\multirow{3}{*}{SU(3)} && \multicolumn{2}{c|}{$T_0 = 0.148, T_c = 0.148$}& \multicolumn{2}{c|}{$T_0 = 0.22, T_c = 0.22$}\\ \cline{3-6}
		& $T_\chi$ &0.198 &0.208&0.218 &0.143 \\ 
		& $T_d$ &0.2 &0.213&0.223&0.163 \\ 
		\hline
	\end{tabular}
	\caption{The parameters $T_0$, $T_c$ and the corresponding pseudo-critical temperature of the chiral phase transition ($T_\chi$), GeV for SU(3) and SU(4) configurations with and without the factor  $F(T) $.}
	\label{tab:Tc2}
\end{table}
We found that the reduced values for $T_0$ are in good agreement with the $N_f$-dependence of $T_0$ suggested in the work \cite{Schaefer:2007pw}. It is also clearly seen that the effect of the U$(1)_A$ anomaly differs between the SU(3) and SU(4) configurations.

\begin{table}[htb]
    \centering
    \begin{tabular}{|c|c|c|c|c|c|}
    \hline
         \multicolumn{3}{|c|}{Pseudoscalar sector ($J^{P}=0^{-}$)}& \multicolumn{3}{|c|}{Scalar sector ($J^{P}=0^{+}$)}\\
    \hline
       Exp., GeV  & SU(3), GeV & SU(4), GeV & Exp., GeV & SU(3), GeV& SU(4), GeV  \\
    \hline
       $\pi$ (0.139)  & 0.139 & 0.142 & $a_0$ (0.98) & 0.926& 0.89  \\
       K (0.497)  & 0.42 & 0.493 & $K_0^*$ (0.63-1.425) & 1.032& 0.993  \\
      $\eta$ (0.547)  & 0.531 & 0.533 & $\sigma$ (0.5-1.2) & 0.7& 0.7 \\
       $\eta^{\prime} (0.958)$  & 0.965 & 0.966 & $f_0$ (0.98-1.5) & 1.124& 1.091  \\
    \hline
    \multicolumn{3}{|c|}{Vector mesons ($J^{P}=1^{-}$)}& \multicolumn{3}{|c|}{Axial-vector mesons ($J^{P}=1^{+}$)} \\
    \hline
    $\rho$ (0.761)  & 0.77 & 0.749 & $a_1$ (1.26) & 1.067 & 1.015   \\
    $K^*$ (0.89)  & 0.851 & 0.903 & $K_1$ (1.27) & 1.253 & 1.276   \\
    $\phi$ (1.02)  & 1.019 & 1.015 & $f_1$ (1.428) & 1.425 & 1.46   \\
    \hline
    \multicolumn{6}{|c|}{Charmed mesons} \\
    \hline
    \multicolumn{3}{|c|}{Scalar-pseudoscalar  sector}  & \multicolumn{3}{|c|}{Vector,axial-vector sector} \\
    \hline
    Exp., GeV  & $J^p$ & SU(4), GeV & Exp., GeV &  $J^p$ & SU(4), GeV   \\
    \hline
    $\eta_c$ (2S) (3.637)  & $0^{-}$ & 4.844 & $J/\psi$ (3.096) & $1^{-}$& 2.403   \\
    $D$ (1.864)  & $0^{-}$ & 2.03 & $D^*$ (2.007) & $1^{-}$& 2.114   \\
    $D_s$ (1.968)  & $0^{-}$ & 2.116 & $D^*_s$ (2.112) & $1^{-}$& 2.1   \\
    $\chi_{c0}$  (3.86) & $0^{+}$ & 3.551 & $\chi_{c1}$ (3.414) & $1^{+}$& 2.937   \\
    $D^*_0$ (2.343)  & $0^{+}$ & 2.578 & $D_1$ (2.422)& $1^{+}$& 2.615   \\
    $D^*_{s0}$ (2.317)  & $0^{+}$ & 2.293 & $D_{s1}$  (2.45) & $1^{+}$& 2.760   \\
    \hline
    \end{tabular}
    \caption{A comparison between meson masses within SU(3) and SU(4) eLSM.}
    \label{tab:mesonMasses}
\end{table}

The various meson states including pseudoscalars ($J^{pc}=0^{-+}$), scalars ($J^{pc}=0^{++}$), vectors ($J^{pc}=1^{--}$), and axial-vectors ($J^{pc}=1^{++}$) are investigated in SU($3$) and SU($4$) eLSM both in vacuum and finite temperatures. The vacuum masses of the mesons are listed in Tab. \ref{tab:mesonMasses} for both configurations and compared with experimental values \cite{ParticleDataGroup:2024cfk}. It is obvious that the masses involving open charm in the scalar and pseudoscalar sectors agree better with experimental values than those in the vector sector. However, the model parameters obtained here do not allow for equally good agreement for quarkonium states.

The examination of the chiral phase-structure properties for all of these mesonic states allows one to determine the in-medium modifications of hadronic matter as well as the critical temperature at which each mesonic state splits into its free quarks. This facilitates a comparison between the masses of strange and charm mesons relative to the masses of lighter mesons. The comparative  analyses in SU($3$) and SU($4$) eLSM helps in determining the influences that brought about additional degrees of freedom and thereby the accurate estimation of the various meson states. 

We conclude that the inclusion of heavy quark in SU($4$), improves the precision of meson mass simulations compared with the SU($3$). The temperature behavior of meson masses appears to show a striking change in the critical temperature region. Our results indicate that different mesonic states display a qualitatively similar behavior near the dissociation region, although the corresponding dissociation temperatures depend on the particular meson state. In addition, we observe that quarkonium states, composed of a heavy quark-antiquark pair, are affected by temperature only gradually, indicating a higher stability against thermal effects.

\section*{Conflicts of Interest}

The authors declare that there are no conflicts of interest regarding the publication of this published article!

\section*{Dataset Availability}

All data generated or analyzed during this study are included in this published article. All of the material is owned by the authors.

\section*{Author contributions}

The responsibility for proposing the conception of the present study lies with AT, who also undertook the tasks of designing and managing the research, interpreting the results, and preparing the manuscript. AF and YuK were responsible for deriving the expressions and proposing physical interpretations. SOA, NMR, and AAA and contributed to the writing and proofreading of the manuscript. The final version of the manuscript was unanimously approved by all authors.

\section*{Funding}

The authors declare that this research received no specific grants from any funding agency in the public, commercial, or not-for-profit sectors.

\section*{Competing interests}

The authors confirm that there are no relevant financial or non-financial competing interests to report.

\bibliographystyle{unsrtnat}
\bibliography{Azar-CharmedMesonStates1-edited}

\newpage 

\appendix
\renewcommand{\thefigure}{A.\arabic{figure}} 
\setcounter{figure}{0} 
\setcounter{table}{0} 
\renewcommand{\thetable}{A\arabic{table}} 

\section{Tree-Level masses} 
\label{sec:App1}

Let us now consider the scalar-pseudoscalar term of the Lagrangian equation (\ref{eq:scalar_nonets}) that includes the determinant term $c[\rm{det}\Phi + \rm{det}\bar{\Phi}]$ as shown in Eq. (\ref{eq:LagrSum}) with $c_0=c_1=0$. If we consider that $\sigma_0$ and $\sigma_8$ fields have nonzero vacuum expectation values ($\bar{\sigma}_0$ and $\bar{\sigma_8}$), and we adjust these fields according to their expectation values \cite{Lenaghan:2000ey}, then the tree-level potential of the LSM in a general form can be expressed as
\begin{eqnarray}
	U(\bar{\sigma})& =& 
	\frac{m^2}{2}\bar{\sigma}_a^2
	-\left[ 
	3 \delta(N_f,2){\cal G}_{ab}+\delta(N_f,3){\cal G}_{abc}\bar{\sigma}_{c}
	\right] \bar{\sigma}_a \bar{\sigma}_b \nonumber\\
	&& + \frac{1}{3} \left[ 
	{\cal F}_{abcd}+\delta(N_f,4){\cal G}_{abcd}
	\right] \bar{\sigma}_a \bar{\sigma}_b \bar{\sigma}_c \bar{\sigma}_d 
	- h_a \bar{\sigma}_a.
	\label{eq:GenMesPot}
\end{eqnarray}
 For the sake of simplicity, the bar notation will be omitted hereafter. The constants ${\cal G}_{ab}$, ${\cal G}_{abc}$, ${\cal G}_{abcd}$, ${\cal F}_{abcd}$, ${\cal H}_{abcd}$ are defined in the following way
\begin{eqnarray}
	{\cal G}_{ab} &=& \frac{c}{6} 
	\left[ 
	\delta_{a0}\delta_{b0} 
	-\delta_{a1}\delta_{b1} 
	-\delta_{a2}\delta_{b2} 
	-\delta_{a3}\delta_{b3} 
	\right], \\
	{\cal G}_{abc} &=&
	\frac{c}{6}
	\bigl[ d_{abc} -
	\frac{3}{2}
	\left( \delta_{a0}d_{0bc} + \delta_{b0}d_{a0c} 
	+ \delta_{c0}d_{ab0} \right)
	+
	\frac{9}{2}
	d_{000}\delta_{a0}\delta_{b0}\delta_{c0}\bigr],\\
	{\cal G}_{abcd} &=& \frac{c}{16} 
	\bigl[ 
	\delta_{ab} \delta_{cd} 
	+\delta_{ad} \delta_{bc}
	+\delta_{ac} \delta_{bd}- 
	\left( 
	d_{abn}d_{ncd} +d_{adn}d_{nbc}+d_{acn}d_{nbd}
	\right)  + 16 \delta_{a0}  \delta_{b0}  \delta_{c0}\delta_{d0}  
	\bigr] \nn \\
	&& - 4 
	\left(
	\delta_{a0}  \delta_{b0}  \delta_{cd}+
	\delta_{a0}  \delta_{c0}  \delta_{bd}+
	\delta_{a0}  \delta_{d0}  \delta_{bc}+
	\delta_{b0}  \delta_{c0}  \delta_{ad}+
	\delta_{b0}  \delta_{d0}  \delta_{ac}+
	\delta_{c0}  \delta_{d0}  \delta_{ab}
	\right) \nn \\ &&
	+ \sqrt{8}
	\left(
	\delta_{a0}d_{bcd}+ \delta_{b0}d_{cda}+\delta_{c0}d_{dab}+\delta_{d0}d_{abc}
	\right), \\
	{\cal F}_{abcd} &=&
	\frac{\lambda_1}{4}
	\left( \delta_{ab}\delta_{cd} 
	+ \delta_{ad}\delta_{bc} + \delta_{ac}\delta_{bd} \right)
	+
	\frac{\lambda_2}{8}
	\left( d_{abn}d_{ncd} + d_{adn}d_{nbc} 
	+ d_{acn}d_{nbd} \right) ,\\
	{\cal H}_{abcd} &=&
	\frac{\lambda_1}{4}
	\delta_{ab}\delta_{cd} +
	\frac{\lambda_2}{8}
	\left( d_{abn}d_{ncd} + f_{acn}f_{nbd}
	+ f_{bcn}f_{nad}\right).
\end{eqnarray}
The standard antisymmetric and symmetric structure constants of SU($N$) are represented by $f_{abc}$ and $d_{abc}$ respectively, where $a,b,c=1,\ldots,N_c^2-1$, with $N_c$ being the degrees of freedom for colors. 

The tree-level masses of pseudoscalars $(m^2_P)_{ab}$ and scalars $(m^2_S)_{ab}$ are determined from the quadratic components of the Lagrangian and, in the general SU($2$)-SU($4$) configurations, can be expressed as
\begin{eqnarray}
	\left( m^2_S \right)_{ab} &=& 
	m^2 \delta_{ab} 
	- 6 \left[\delta(N_f, 2) {\cal G}_{ab}+\delta(N_f,3){\cal G}_{abc} \bar{\sigma}_c\right] + 4 \left[
	{\cal F}_{abcd} +\delta(N_f, 4){\cal G}_{abcd}\right]\bar{\sigma}_c \bar{\sigma}_d,
	\\
	\left( m^2_P \right)_{ab} &=& m^2 \delta_{ab} 
	+ 6 \left[\delta(N_f, 2) {\cal G}_{ab}+\delta(N_f,3){\cal G}_{abc} \bar{\sigma}_c\right] + 4 \left[
	{\cal F}_{abcd} -\delta(N_f, 4){\cal G}_{abcd}\right]\bar{\sigma}_c \bar{\sigma}_d  .
\end{eqnarray}
Here, the summation runs over the index $n$ only. Equations for vector and axial-vector mesons are defined as 
\begin{eqnarray}
	(m_V^2)_{ab} &=& m_1^2+J_{abmn}\sigma_{m} \sigma_{n}, \\
	(m_A^2)_{ab} &=& m_1^2+J'_{abmn}\sigma_{m} \sigma_{n},
\end{eqnarray}
where $a, b$ refer to the meson state, and the summation runs over $m, n \in \{1,2,\cdots,(N^2_c-1)$  \cite{Parganlija:2012fy}. The coefficients $J_{abcd}$ and $J'_{abcd}$ are given as follows 
\begin{eqnarray}
	J_{abcd}&=& g_1^2f_{acn}f_{bdn} + \frac{h_1}{2}\delta_{ab}\delta_{cd} + \frac{h_2}{2}d_{abn}d_{cdn} \nn \\
	&+& \frac{h_3}{2}\left(d_{acn}d_{bdn}+d_{adn}d_{bcn}-f_{acn}f_{bdn}-f_{adn}f_{bcn}\right),\\
	J'_{abcd}&=& g_1^2d_{acn}d_{bdn} + \frac{h_1}{2}\delta_{ab}\delta_{cd} + \frac{h_2}{2}d_{abn}d_{cdn} \nn \\
	&-& \frac{h_3}{2}\left(d_{acn}d_{bdn}+d_{adn}d_{bcn}-f_{acn}f_{bdn}-f_{adn}f_{bcn}\right).
\end{eqnarray}
Again, the sum operates solely over the index $n$, while $d_{abc}$ and $f_{abc}$ represent the symmetric and antisymmetric structure constants of U($N_f$), respectively. It should be noticed that all the resulting expressions are converted to the nonstrange-strange and the nonstrange-strange-charm basis as it was discussed in Section \ref{sec:su3} and Section \ref{sec:su4}.

Furthermore, it should be emphasized that due to the mixing between the (pseudo)scalar and (axial)vector sector through the covariant derivative, Eq. (\ref{eq:covarD}), the tree-level expressions of the pseudoscalars and some scalars are not mass eigenstates. It was reported that when the corresponding wave functions are renormalized to the constants $Z_i$, such a mixing can be resolved  \cite{Parganlija:2012gv}. The factors $Z_i$ are now emerging in mass expressions \cite{Tawfik:2025hhr}
\begin{eqnarray}
	Z_{\pi} &=& Z_{\eta_N} = \frac{m_{a_1}}{\sqrt{m^2_{a_1}-g^2_1 \sigma^2_x}},  \qquad  \qquad
	Z_K= \frac{2 m_{K_1}}{\sqrt{4 m^2_{K_1} - g^2_1\left(\sigma_x + \sqrt{2} \sigma_y\right)^2}}, \nn \\
	Z_{\eta_S} &=& \frac{m_{f_{1S}}}{\sqrt{m^2_{f_{1S}} - 2 g^2_1 \sigma^2_y}},\qquad  \qquad \qquad\;
	Z_{K^{\ast}_0} = \frac{2 m_{K^{\ast}}}{\sqrt{4 m^2_{K^{\ast}} - g^2_1 \left(\sigma_x - \sqrt{2} \sigma_y\right)^2}}, 
\end{eqnarray}
And for charmed mesons, the factors $Z_i$ read \cite{Tawfik:2025hhr}
\begin{eqnarray}
	Z_{\eta_c} &=& \frac{m_{\chi_{c1}}}{\sqrt{m_{\chi_{c1}}^2 - 2 g_1^2 \sigma_c^2}}, \qquad  \qquad \qquad \quad
	Z_D= \frac{2 m_{D_1}}{\sqrt{4 m_{D_1} - g_1^2 \left(\sigma_x + \sqrt{2}\sigma_c\right)^2}}, \nn \\
	Z_{D_s} &=& \frac{\sqrt{2} m_{D_{s_1}}}{\sqrt{2 m_{D_{s_1}}^2- g_1^2 \left(\sigma_y + \sigma_c\right)^2}}, \quad  \qquad
	Z_{D_0^*}= \frac{2 m_{D^*}}{\sqrt{4 m_{D^*} - g_1^2 \left(\sigma_x - \sqrt{2}\sigma_c\right)^2}}, \nn \\
	Z_{D_0^{*0}} &=& \frac{2 m_{D^{*0}}}{\sqrt{4 m_{D^{*0}}^2 - g_1^2 \left(\sigma_x - \sqrt{2}\sigma_c\right)^2}}, \quad \;\;
	Z_{D_{s0}^{*}}= \frac{ \sqrt{2} m_{D^{*}_{s}} }{
		\sqrt{2 m_{D^{*}_{s}}^2 - g_1^2 \left(\sigma_y - \sigma_c\right)^2}
	}. \nn
\end{eqnarray}

\section{Meson states}
\label{sec:AppB}

For the $N_f=3$ the identification of the physical scalar and pseudoscalar  fields is given as
\begin{eqnarray}
	T_a \sigma_a &=& \frac{1}{\sqrt{2}}\left( 
	\begin{array}{ccc}
		\frac{1}{\sqrt{2}} a_0^0 + \frac{1}{\sqrt{6}}\sigma_8 + \frac{1}{\sqrt{3}}\sigma_0 & a_0^+ & K^{*+}_0 \\
		a_0^- &  - \frac{1}{\sqrt{2}} a_0^0 + \frac{1}{\sqrt{3}}\sigma_0+ \frac{1}{\sqrt{6}}\sigma_8 & K^{*0}_0 \\
		{K}^{*-}_0 & \bar{K}^{*0}_0 &    - \frac{2}{\sqrt{6}} \sigma_8 +\frac{1}{\sqrt{3}}\sigma_0 
	\end{array}
	\right),  \\
	T_a \pi_a &=& \frac{1}{\sqrt{2}}\left( 
	\begin{array}{ccc}
		\frac{1}{\sqrt{2}} \pi^0 +\frac{1}{\sqrt{6}} \pi_8+ \frac{1}{\sqrt{3}}\pi_0 & \pi^+ & K^+\\
		\pi^- &  - \frac{1}{\sqrt{2}} \pi^0 +\frac{1}{\sqrt{6}} \pi_8+\frac{1}{\sqrt{3}}\pi_0 & K^- \\
		K^-   & \bar{K}^0 &  - \frac{2}{\sqrt{6}} \pi_8 +\frac{1}{\sqrt{3}}\pi_0  
	\end{array}
	\right).
\end{eqnarray}
Here $\pi^{\pm}=(\pi_1\mp i\pi_2)/\sqrt{2}$ and $\pi^0=\pi_3$ are the pions,  $K^{\pm}=(\pi_4\mp i\pi_5)/\sqrt{2},K^0=(\pi_6- i\pi_7)/\sqrt{2}, \bar{K}^0=(\pi_6+ i\pi_7)/\sqrt{2}$ are the kaons. The $\pi_0$ and the $\pi_8$  are admixtures of the $\eta$ and $\eta'$ mesons and the fields $\sigma_0$ and $\sigma_8$ correspond to admixtures of the $\sigma$ and $f_0$ mesons. 

For $N_f=4$ we use the following matrices
\begin{eqnarray}
\begin{array}{ccl}
	T_a \sigma_a = \frac{1}{\sqrt{2}}\left( 
	\begin{array}{cccc}
		A_S     & a_0^+          & K^{*+}_0  & \bar{D}_0^0  \\
		a_0^-    &  B_S           & K^{*0}_0  &  \bar{D}_0^- \\
		{K}^{*-}_0 & \bar{K}^{*0}_0 &  C_S      &  D_{s0}^-    \\
		{D}_0^0 & {D}_0^+        &  D_{s0}^+ & D_S 
	\end{array}
	\right)  , \qquad
	T_a \pi_a = \frac{1}{\sqrt{2}}\left( 
	\begin{array}{cccc}
		A_P   & \pi^+      & K^+   &  {D}^0 \\
		\pi^- &  B_P       & K^0   &  {D}^-\\
		K^-   & \bar{K}^0 &  C_P   &  D_{s}^-    \\
		{D}^0 & {D}^+        &  D_{s}^+ & D_P 
	\end{array}
	\right), 
	\end{array}
\end{eqnarray}
where the diagonal elements are given as
\begin{eqnarray}
	\begin{array}{ccl}
		A_S & = & \frac{1}{2}\sigma_0 +\frac{1}{\sqrt{2}} a_0^0 + \frac{1}{\sqrt{6}}\sigma_8 
		+\frac{1}{\sqrt{12}}\sigma_{15}, \\ 
		B_S & = & \frac{1}{2}\sigma_0 -\frac{1}{\sqrt{2}}a_0^0 + \frac{1}{\sqrt{6}}\sigma_8 
		+\frac{1}{\sqrt{12}}\sigma_{15}, \\ 
		C_S & = & \frac{1}{2}\sigma_0                           - \frac{2}{\sqrt{6}}\sigma_8 
		+\frac{1}{\sqrt{12}}\sigma_{15}, \\ 
		D_S & = & \frac{1}{2}\sigma_0 
		-\frac{3}{\sqrt{12}}\sigma_{15}, \\ 
	\end{array} \,\,\, 
	\begin{array}{ccl}
		A_P & = & \frac{1}{2}\pi_0 +\frac{1}{\sqrt{2}}\pi^0 + \frac{1}{\sqrt{6}}\pi_8 
		+\frac{1}{\sqrt{12}}\pi_{15}, \\ 
		B_P & = & \frac{1}{2}\pi_0 -\frac{1}{\sqrt{2}}\pi^0 + \frac{1}{\sqrt{6}}\pi_8 
		+\frac{1}{\sqrt{12}}\pi_{15}, \\ 
		C_P & = & \frac{1}{2}\pi_0                           - \frac{2}{\sqrt{6}}\pi_8 
		+\frac{1}{\sqrt{12}}\pi_{15}, \\ 
		D_P & = & \frac{1}{2}\pi_0 
		-\frac{3}{\sqrt{12}}\pi_{15}. \\ 
	\end{array} 
\end{eqnarray}
The nondiagonal pseudoscalar elements are defined as:
\bea 
\begin{array}{ccl}
D^0 &=& \frac{1}{\sqrt{2}}(\pi_9+i\pi_{10}), \\ 
\bar{D}^0&=& \frac{1}{\sqrt{2}}(\pi_9-i\pi_{10}), \\ D^\pm &=& \frac{1}{\sqrt{2}} (\pi_{11} \pm i \pi_{12}), \\  
D_s^\pm &=& \frac{1}{\sqrt{2}} (\pi_{13} \pm i \pi_{14}). 
\end{array}
\eea  
The nondiagonal scalar mesons are represented as
\bea 
\begin{array}{ccl}
D_0^0 &=& \frac{1}{\sqrt{2}} (\sigma_9+i\sigma_{10}), \\
\bar{D}_0^0 &=& \frac{1}{\sqrt{2}} (\sigma_9-i\sigma_{10}),  \\ 
D_0^\pm &=& \frac{1}{\sqrt{2}} (\sigma_{11} \pm i \sigma_{12}), \\  
D_{s0}^\pm &=& \frac{1}{\sqrt{2}}(\sigma_{13} \pm i \sigma_{14}).
\end{array}
\eea

The mixing of isoscalar states in pseudoscalar and scalar SU($3$) multiplets is described in detail in Refs. \cite{Parganlija:2012gv,Schaefer:2008hk}. The $\pi_0, \pi_8$ and $\pi_{15}$ fields are admixtures of the $\eta,\eta', \eta_c$ mesons and the scalar fields $\sigma_0, \sigma_8$ and $\sigma_{15}$  are admixtures of the $\sigma, f_0$ and $\chi_{c0}$ mesons. In SU($4$) eLSM, the interpretation of mixing pattern is assumed to be similar to that of its SU($3$) configuration. Accordingly, the isoscalar states in the 15-plet -singlet  ($\eta_0, \eta_8, \eta_{15}$) are defined as
\begin{eqnarray}
\begin{array}{ccl}
	|\eta_0\rangle& =& \frac{1}{2}(|u\bar{u}\rangle + |d\bar{d}\rangle +|s\bar{s}\rangle+|c\bar{c}\rangle),\\
	|\eta_8\rangle& =& \frac{1}{\sqrt{6}}(|u\bar{u}\rangle + |d\bar{d}\rangle -2 |s\bar{s}\rangle),\\
	|\eta_{15}\rangle& =& \frac{1}{\sqrt{12}}(|u\bar{u}\rangle + |d\bar{d}\rangle +|s\bar{s}\rangle -3|c\bar{c}\rangle).
\end{array}
\end{eqnarray}
Defining the eigenstates for the nonstrange-strange-charm basis as $\eta_{N} = (|u\bar{u}\rangle+|d\bar{d}\rangle)/\sqrt{2} $, $\eta_S = |s\bar{s}\rangle$, $\eta_{c}=|c\bar{c}\rangle$,  the transition matrix for SU(4) can be written as:
\begin{eqnarray}\label{eq:eta_base}
	\left( 
	\begin{array}{c}
		\eta_N   \\
		\eta_S   \\
		\eta_C   
	\end{array}
	\right)   = 
	\left( 
	\begin{array}{ccc}
		\frac{1}{\sqrt{2}}    & \frac{1}{\sqrt{3}} & \frac{1}{\sqrt{6}}   \\
		\frac{1}{2}    &  -\sqrt{\frac{2}{3}}           & \frac{1}{2\sqrt{3}}  \\
		\frac{1}{2} & 0 & -\frac{\sqrt{3}}{2}         \\
	\end{array}
	\right) 
	\left( 
	\begin{array}{c}
		\eta_0       \\
		\eta_8    \\
		\eta_{15}     
	\end{array}
	\right).
\end{eqnarray}
This full $3 \times 3$ mixing matrix can be considered as block-diagonal, specifically, a $2 \times2$ block for the light $\eta/ \eta'$ system and a $1 \times 1$ block for the heavy $\eta_c$ \cite{Feldmann:1998vh}.  The physical particles can be realized in the same way as in SU(3) $|\eta\rangle\approx \cos\theta_P |\eta_8\rangle - \sin\theta_P |\eta_0\rangle$, $|\eta'\rangle\approx \sin\theta_P |\eta_8\rangle +\cos\theta_P |\eta_0\rangle$, where the mixing angle $\theta_P$ is experimentally found to be about $-41^{\circ}$ \cite{DiDonato:2011kr,Klempt:2004yz}. Using the basis in Eq. \eqref{eq:eta_base}, the mass elements for $\eta$ mesons can be written as \cite{Schaefer:2008hk}.
\begin{equation}
	m^2_{\eta'/\eta} = \frac{1}{2}\left(m^2_{\eta_N}+m^2_{\eta_S}\pm\sqrt{\left(m^2_{\eta_N}-m^2_{\eta_S}\right)^2-4 m^4_{\eta_{NS}}} \right),
\end{equation}
with $m_{\eta_N}$, $m_{\eta_S}$, $m_{\eta_{NS}}$ defined in Tables below.  We keep the same structure for physical particles as in SU(3) because of the heavy mass of the charm quark. It is assumed to be decoupled almost completely from the light quarks. This means that the off-diagonal elements $\eta_{8}\eta_{15}$ and $\eta_0\eta_{15}$ become relatively small relative to $\eta_{15}\eta_{15}$ and can be neglected also due to the mixing angles $\theta_{015}$ and $\theta_{815}$ are small \cite{Feldmann:1998vh,Bagchi:1999dx}. It should be noted that the state $\eta_{15}$ can be identified the charmonium state $\eta_{c}(1S)$ \cite{Navelet:1980fp, Zhang:2025yeu,Feldmann:1998vh}.

For the scalar sector, considering the light sector only (ignoring charm completely), the mixing among the three relevant states can be parametrized in the same way as discussed above with corresponding replacements ($\sigma_0, \sigma_8$), ($\sigma_N, \sigma_S$) and the scalar mixing angle $\theta_S$ \cite{Schaefer:2008hk}. 
\begin{equation}
	m^2_{\sigma/f_0} = \frac{1}{2}\left(m^2_{\sigma_N}+m^2_{\sigma_S}\pm\sqrt{\left(m^2_{\sigma_N}-m^2_{\sigma_S}\right)^2-4 m^4_{\sigma_{NS}}} \right).
\end{equation}
The vector $V^\mu$ and the axial-vector $A^\mu$ sectors are described by the following matrices:
\begin{eqnarray}
V^\mu &=& \frac{1}{\sqrt{2}}\left( 
	\begin{array}{cccc}
		\frac{1}{\sqrt{2}}(\omega_N+\rho^0)     & \rho^+          & K^{*+}  & {D}^{*0}  \\
		\rho^-    &  \frac{1}{\sqrt{2}}(\omega_N-\rho^0)           & K^{*0}  &  D^{*-} \\
		K^{*-} &\bar{K}^{*0} & \omega_S      &  D_{s}^{*-}    \\
		\bar{D}^{*0} & {D}^{*+}        &  D_{s}^{*+} & J/\psi
	\end{array}
	\right), \\ 
A^{\mu} &=& \frac{1}{\sqrt{2}}\left( 
	\begin{array}{cccc}
		\frac{1}{\sqrt{2}}(f_{1,N}+a_1^0)     & a_1^+          & K_1^+  & \bar{D}_1^0  \\
		a_1^-    &  \frac{1}{\sqrt{2}}(f_{1,N}-a_1^0)           & K_1^{0}  &  D_1^{-} \\
		K_1^{-} & \bar{K}_1^{0} & f_{1S}      &  D_{s1}^{-}    \\
		\bar{D}_1^0 & D_1^{+}        &  D_{s1}^{+} & \chi_{c1}
	\end{array}
	\right). 
\end{eqnarray}
The missing between strange and non-strange isoscalars is neglected here \cite{Eshraim:2014eka,Klempt:2004yz}. 

The mass expressions obtained for the listed states are presented in Tables \ref{tab:su3masses} - \ref{tab:su4charmed} below. As it was said above, we express all meson masses in term of the strange--non-strange--(charm) basises.

\begin{table}[h]
\Large 
	\centering
\resizebox{\textwidth}{!}{
	\begin{tabular}{|c|c|}
		\hline
		Scalar mesons & Pseudoscalar mesons \\
		\hline
		\vbox{
			\begin{eqnarray*}
				m^2_{a_0} &=& m^2_0 + \frac{c}{\sqrt{6}}\sigma_y+ \left(\lambda_1 + \frac{3}{2} \lambda_2\right) \sigma^2_x + \lambda_1\sigma_y,\\
				m^2_{K_0^{\ast}} &=& Z^2_{K_0^{\ast}} \\
				&&\times\left(m^2_0 + \frac{c}{\sqrt{2}}\sigma_y + \left(\lambda_1+\frac{1}{2}\lambda_2\right) \sigma^2_x + \frac{\lambda_2}{\sqrt{2}}\sigma_x\sigma_y  +\left(\lambda_1+\lambda_2\right) \sigma_y^2 \right), \\
				m^2_{\sigma_N} &=&m^2_0 -\frac{c \sigma_y}{\sqrt{2}}+\frac{3}{2}(2 \lambda_1+\lambda_2) \sigma_x^2 +\lambda_1 \sigma_y^2, \\
				m^2_{\sigma_S} &=& m^2_0 + 3 (\lambda_1+\lambda_2) \sigma_y^2 + \lambda_1 \sigma_x^2, \\
				m^2_{\sigma_{SN}}&=& -\frac{c \sigma_x}{\sqrt{2}} + 2 \lambda_1 \sigma_x \sigma_y.
			\end{eqnarray*}
		} &
		\vbox{
			\begin{eqnarray*}
				m^2_{\pi} &=& Z^2_{\pi}\left( m^2_0 - \frac{c}{\sqrt{2}}\sigma_y + \left(\lambda_1 + \frac{\lambda_2}{2}\right) \sigma^2_x + \lambda_1\sigma^2_y \right), \\
				m^2_{K}&=& Z^2_{K} \\
				&&\times\left(m^2_0 - \frac{c}{2}\sigma_x + \left(\lambda_1+\frac{\lambda_2}{2}\right) \sigma_x^2 + (\lambda_1 +\lambda_2)\sigma_y^2-\frac{\lambda_2}{\sqrt{2}} \sigma_x \sigma_y\right), \\
				m^2_{\eta_N} &=& Z^2_\pi \left(m_0^2 + \frac{c}{\sqrt{2}}\sigma_y+\left(\lambda_1 + \frac{\lambda_2}{2}\right)  \sigma_x^2 + \lambda_1\sigma_y^2\right)\\
				m^2_{\eta_S}&=& Z^2_{\eta_S} \left(m_0^2 + \lambda_1 \sigma_x^2 + (\lambda_1 + \lambda_2) \sigma_y^2\right)\\
				m^2_{\eta_{NS}} &=& Z^2_\pi Z^2_{\eta_S} \frac{c}{2}\sigma_x.
			\end{eqnarray*}
		}  \\ 
		\hline
		Vector mesons & Axial-vector mesons\\
		\hline
		\vbox{
			\begin{eqnarray*}
				m^2_{\rho} &=& m^2_1 + \frac{h_1}{2}(\sigma_x^2+\sigma_y^2) +\frac{1}{2}\left(h_2+h_3\right) \sigma_x^2 + 2 \delta_x,\\
				m^2_{K^{\ast}} &=& m^2_1+ \frac{h_1}{2}(\sigma_x^2+\sigma_y^2) + \frac{1}{4}\left(g_1^2 + h_2\right)\left(\sigma_x^2 + 2 \sigma_y^2\right)\\
				&& + \frac{\sigma_x \sigma_y}{\sqrt{2}}\left(h_3-g_1^2\right) + \delta_x + \delta_y,\\
				m^2_{\omega_{S}} &=& m^2_1 + \frac{h_1}{2}(\sigma_x^2+\sigma_y^2)+ \left(h_2 + h_3\right)\sigma_y^2  + 2 \delta_y,  \\
				m^2_{\omega_{N}} &=& m^2_{\rho}.
			\end{eqnarray*}
		}  &
		\vbox{
			\begin{eqnarray*}
				m^2_{a_1} &=& m^2_1 + g_1^2\sigma_x^2 + \frac{h_1}{2}(\sigma_x^2+\sigma_y^2) + \frac{1}{2}(h_2-h_3) \sigma_x^2  + 2 \delta_x,\\
				m^2_{K_1} &=& m^2_1 + \frac{h_1}{2}(\sigma_x^2+\sigma_y^2) +  \frac{1}{4}\left(g_1^2 + h_2\right) \left(\sigma_x^2+2\sigma_y^2\right)\\
				&& - \frac{\sigma_x \sigma_y }{\sqrt{2}}\left(h_3-g_1^2\right) + \delta_x + \delta_y,\\
				m^2_{f_{1S}} &=& m^2_1 + 2 g_1^2\sigma_y^2 + \frac{h_1}{2}(\sigma_x^2+\sigma_y^2)  + \left(h_2 - h_3\right)\sigma_y^2 + 2 \delta_y,  \\
				m^2_{f_{1N}} &=& m^2_{a_1}.
			\end{eqnarray*}
		}\\
		\hline
	\end{tabular}
}
\caption{The meson masses for SU($3$) configuration}
	\label{tab:su3masses}
\end{table}

\begin{table}[h]
\Large 
	\centering
\resizebox{\textwidth}{!}{
	\centering
	\begin{tabular}{|c|c|}
		\hline
		Scalar non-charmed  mesons & Pseudoscalar non-charmed  mesons \\
		\hline
		\vbox{
			\begin{eqnarray*}
				m_{a_{0}}^2 &=& m_{0}^2 + \frac{c}{2}\sigma_c\sigma_y + \lambda_1\left(\sigma_x^2+\sigma_y^2+\sigma_c^2\right) + \frac{3\lambda_2}{2} \sigma_x^2, \\
				m^2_{K_0^{\ast}} &=& Z_{K_0^{\ast}}^2\Big( m^2_0 + \frac{c}{2 \sqrt{2}}\sigma_x\sigma_c + \left(\lambda_1+ \frac{\lambda_2}{2}\right)\sigma_x^2   + \lambda_1\left(\sigma_y^2+\sigma_c^2\right)\\
				&& +  \frac{\lambda_2}{\sqrt{2}}\sigma_x\sigma_y + \lambda_2 \sigma_y^2\Big),  \\
				m_{{\sigma}_{N}}^2 &=& m_{0}^2 - \frac{c}{2}\sigma_c\sigma_y + 3\left(\lambda_1 +\frac{\lambda_2}{2}\right)\sigma_x^2 +\lambda_1\left(\sigma_y^2+\sigma_c^2\right) \\
				m_{{\sigma}_{S}}^2 &=& m_{0}^2 + \lambda_1\left(\sigma_x^2+\sigma_c^2\right) + 3\left(\lambda_1+\lambda_2\right) \sigma_y^2, \\
				m_{{\sigma}_{NS}}^2 &=& 2\lambda_1\sigma_x\sigma_y - \frac{c}{2}\sigma_c\sigma_x.
			\end{eqnarray*}
		}&
		\vbox{
			\begin{eqnarray*}
				m_{\pi}^2 &=& Z_{\pi}^2\left(m_0^2 - \frac{c}{2}\sigma_y\sigma_c + \left(\lambda_1+ \frac{\lambda_2}{2} \right)\sigma_x^2+\lambda_1 \sigma_y^2 + \lambda_1 \sigma_c^2\right), \\
				m_{K}^2 &=& Z_{K}^2 \Big(m_0^2 - \frac{c}{2 \sqrt{2}}\sigma_x\sigma_c + \left(\lambda_1+\frac{\lambda_2}{2}\right)\sigma_x^2 + \lambda_1 \left(\sigma_y^2+\sigma_c^2\right)\\
				&&-\frac{\lambda_2}{\sqrt{2}} \sigma_x \sigma_y + \lambda_2 \sigma_y^2\Big), \\
				m_{{\eta}_{N}}^2 &=& Z_{\pi}^2\left(m_0^2+\frac{c}{2}\sigma_c\sigma_y + \left(\lambda_1+\frac{\lambda_2}{2}\right)\sigma_x^2+\lambda_1 \left(\sigma_y^2+ \sigma_c^2\right) \right), \\
				m_{{\eta}_{S}}^2 &=& Z_{{\eta}_{S}}^2 \left(m_0^2 + \lambda_1\left(\sigma_x^2+\sigma_c^2\right) + \left(\lambda_1+\lambda_2\right) \sigma_y^2 \right),\\
				m_{{\eta}_{NS}}^2 &=& - Z_{\pi}Z_{{\eta}_{S}}\frac{c}{2}\sigma_x\sigma_c.
			\end{eqnarray*}
		}\\
		\hline
		Vector  non-charmed mesons & Axial-vector non-charmed mesons\\
		\hline
		\vbox{
			\begin{eqnarray*}
				m^2_{\rho} &=& m_1^2 + \frac{h_1}{2} \left(\sigma_x^2+\sigma_y^2+\sigma_c^2\right)+\frac{1}{2}\left(h_2+h_3\right)\sigma_x^2 + 2 \delta_x, \\
				m^2_{K^{\ast}} &=& m_1^2 +  \frac{h_1}{2} \left(\sigma_x^2+\sigma_y^2+\sigma_c^2\right) +  \frac{1}{4}  \left(g_1^2 +  h_2\right)\left(\sigma_x^2+2 \sigma_y^2\right) \\
				&& +\frac{\sigma_x\sigma_y}{\sqrt{2}} \left(h_3 - g_1^2\right)  + \delta_x + \delta_y, \\
				m^2_{\omega_S} &=& m_1^2 + \frac{h_1}{2} \left(\sigma_x^2 + \sigma_y^2+ \sigma_c^2\right) + \left(h_2+h_3\right) \sigma_y^2 +  2 \delta_y, \\
				m^2_{\omega_N}&=& m^2_{\rho}.
			\end{eqnarray*}
		}&
		\vbox{
			\begin{eqnarray*}
				m^2_{a_1} &=& m_1^2 + g_1^2\sigma_x^2+ \frac{h_1}{2} \left(\sigma_x^2+\sigma_y^2+\sigma_c^2\right)+\frac{1}{2}\left(h_2-h_3\right)\sigma_x^2 + 2 \delta_x, \\
				m^2_{K_1} &=& m_1^2 +  \frac{h_1}{2} \left(\sigma_x^2+\sigma_y^2+\sigma_c^2\right) +  \frac{1}{4}  \left(g_1^2 +  h_2\right)\left(\sigma_x^2+2 \sigma_y^2\right) \\
				&& -\frac{\sigma_x\sigma_y}{\sqrt{2}} \left(h_3 - g_1^2\right)  + \delta_x + \delta_y, \\
				m^2_{f_{1S}} &=& m_1^2 + 2 g_1^2\sigma_y^2 + \frac{h_1}{2} \left(\sigma_x^2 + \sigma_y^2+ \sigma_c^2\right) + \left(h_2-h_3\right) \sigma_y^2 +  2 \delta_y, \\
				m^2_{f_{1N}}&=& m^2_{a_1}.
			\end{eqnarray*}
		}\\
		\hline
	\end{tabular}
}
	\caption{The non-charmed meson masses for SU($4$) configuration.}
	\label{tab:su4noncahrmed}
\end{table}

\begin{table}[h]
\Large 
	\centering
\resizebox{\textwidth}{!}{
	\centering
	\begin{tabular}{|c|c|}
		\hline
		Scalar charmed  mesons & Pseudoscalar charmed  mesons\\
		\hline
		\vbox{
			\begin{eqnarray*}
				m_{{\chi}_{c0}}^2&=& m_0^2 + \lambda_1 \left(\sigma_x^2+\sigma_y^2\right) + 3\left(\lambda_1 +\lambda_2\right)\sigma_c^2 + 2\epsilon_c, \\
				m_{{D}_{0}^\ast}^2&=& Z_{D_0^*}^2\Big(m^2_0- \frac{c}{2 \sqrt{2}}\sigma_x\sigma_y +\left(\lambda_1 +\frac{\lambda_2}{2}\right) \sigma_x^2+\lambda_1\sigma_y^2 + \left(\lambda_1+\lambda_2\right)\sigma_c^2\\
				&&+ \frac{\lambda_2}{\sqrt{2}}\sigma_x\sigma_c  + \epsilon_c\Big), \\
				m_{{D}_{s0}^{\ast}}^2 &=& Z_{D_{s0}^{\ast}}^2\left(m_0^2 - \frac{c}{4}\sigma_x^2+\lambda_1\sigma_x^2 + \left(\lambda_1+\lambda_2\right) \sigma_y^2 +\lambda_2\sigma_y\sigma_c+\left(\lambda_1+\lambda_2\right)\sigma_c^2 +\epsilon_c\right).
			\end{eqnarray*}
		} &
		\vbox{
			\begin{eqnarray*}
				m_D^2 &=& Z_D^2\Big(m_0^2 + \frac{c}{2\sqrt{2}}\sigma_x\sigma_y +\left(\lambda_1+\frac{\lambda_2}{2}\right)\sigma_x^2 + \lambda_1\sigma_y^2 +\left(\lambda_1+\lambda_2\right)\sigma_c^2 \\
				&& - \frac{\lambda_2}{\sqrt{2}} \sigma_x \sigma_c + \epsilon_c\Big), \\
				m_{\eta_c}^2 &=& Z_{\eta_c}^2\left(m_0^2 +\lambda_1\left(\sigma_x^2 + \sigma_y^2\right)+\left(\lambda_1+\lambda_2\right)\sigma_c^2 + 2\epsilon_c \right), \\
				m_{D_s}^2 &=& Z_{D_s}^2\left(m_0^2 + \frac{c}{4}\sigma_x^2+\lambda_1 \sigma_x^2 + \left(\lambda_1+\lambda_2\right)\sigma_y^2 +\left(\lambda_1+\lambda_2\right)\sigma_c^2 - \lambda_2\sigma_y\sigma_c + \epsilon_c\right).
			\end{eqnarray*}
		}\\
		\hline
		Vector charmed mesons & Axial-vector charmed mesons\\
		\hline
		\vbox{
			\begin{eqnarray*}
			m_{D^{\ast}}^2 &=& m_1^2 + \frac{h_1}{2}\left(\sigma_x^2+\sigma_y^2+\sigma_c^2\right)+\frac{1}{4}\left(g_1^2+h_2\right)(\sigma_x^2+ 2\sigma_c^2)\\ 
				&&+ \frac{\sigma_x\sigma_c}{\sqrt{2}} \left(h_3-g_1^2\right)+\delta_x+\delta_c,  \\
				m_{J/\psi}^2 &=& m_1^2+\frac{h_1}{2} \left(\sigma_x^2+\sigma_y^2+\sigma_c^2\right) +\left(h_2+h_3\right)\sigma_c^2 + 2\delta_c, \\
				m_{D_s^\ast}^2 &=& m_1^2+\frac{h_1}{2}\left(\sigma_x^2+\sigma_y^2+\sigma_c^2\right)+\frac{1}{2}\left(g_1^2+h_2\right) \left(\sigma_y^2+\sigma_c^2\right)\\
				&& + \left(h_3-g_1^2\right)\sigma_y\sigma_c + \delta_y + \delta_c.
			\end{eqnarray*}
		} &
		\vbox{
			\begin{eqnarray*}
				m_{D_1}^2 &=& m_1^2+  \frac{h_1}{2}\left(\sigma_x^2+\sigma_y^2+\sigma_c^2\right)  +\frac{1}{4}\left(g_1^2+h_2\right)\left(\sigma_x^2+2\sigma_c^2\right)\\
				&&+ \frac{\sigma_x\sigma_c}{\sqrt{2}}\left(g_1^2 -h_3\right) +\delta_x + \delta_c.\\
				m_{{\chi}_{c1}}^2 &=& m_1^2+ 2 g_1^2\sigma_c^2+\frac{h_1}{2}\left(\sigma_x^2+\sigma_y^2+\sigma_c^2\right) + (h_2-h_3)\sigma_c^2 + 2\delta_c.\\   
				m_{D_{s1}}^2&=& m_1^2+\frac{h_1}{2}\left(\sigma_x^2+\sigma_y^2+\sigma_c^2\right)+\frac{1}{2}\left(g_1^2+h_2\right) \left(\sigma_y^2+\sigma_c^2\right)\\
				&&+\sigma_y\sigma_c\left(g_1^2-h_3\right) +\delta_y + \delta_c, \\
			\end{eqnarray*}
		}\\
		\hline
	\end{tabular}
}
	\caption{ The masses of mesons with open (hidden) charm.}
	\label{tab:su4charmed}
\end{table}

\clearpage

\section{Derivations of Squared Quark Mass}
\label{appndC}
In Section \ref{sec:finiteT}, it was noted that the in-medium meson masses are influenced by the quark contribution, Eq. \eqref{eq:ftmass}. The results for the first and second derivatives of the quark mass with respect to the meson fields for the SU($3$) configuration can be found for example in Refs. \cite{Schaefer:2008hk,Gupta:2009fg} and in presence of (axial-)vector fields in \cite{Kovacs:2016juc}. Our results for  SU($4$) configuration are presented in Tab. \ref{tab:deriv}. The degenerate light flavours are presented in second and third, $m_{f,a}^2m_{f,b}^2/g^4$, $m_{f,ab}^2/g^2$, the strange quark in fourth and fifth, $m_{y,a}^2m_{y,b}^2/g^4$, $m_{y,ab}^2/g^2$, and the charm quark in last two columns,  $m_{c,a}^2m_{c,b}^2/g^4$, $m_{c,ab}^2/g^2$. 
\begin{table}[h]
	\centering
	\begin{tabular}{|c|c|c|c|c|c|c|}
		\hline
		&   $m_{f,a}^2m_{f,b}^2/g^4$& $m_{f,ab}^2/g^2$& $m_{y,a}^2m_{y,b}^2/g^4$& $m_{y,ab}^2/g^2$& $m_{c,a}^2m_{c,b}^2/g^4$&$m_{c,ab}^2/g^2$\\
		\hline
		$\sigma_0 \ \sigma_0$& 
		$\frac{1}{4}\sigma_x^2$& $\frac{1}{2}$& $\frac{1}{4}\sigma_y^2$& $\frac{1}{4}$& $\frac{1}{4}\sigma_c^2$&$\frac{1}{4}$\\
		$\sigma_1 \ \sigma_1$& 0& 1& 0& 0& 0&0\\
		$\sigma_4 \ \sigma_4$& 0& $\frac{Z_{K_0^*}^2\sigma_x}{\sigma_x -\sqrt{2} \sigma_y}$& 0&$\frac{Z_{K_0^*}^2\sqrt{2}\sigma_y}{\sqrt{2}\sigma_y -\sigma_x}$& 0&0\\
		$\sigma_9 \ \sigma_9$& 0& $\frac{Z_{D_0^*}^2\sigma_x}{\sigma_x-\sqrt{2} \sigma_c}$& 0& 0& 0&$\frac{Z_{D_0^*}^2\sqrt{2}\sigma_c}{\sqrt{2} \sigma_c- \sigma_x}$\\
		$\sigma_{13} \ \sigma_{13}$& 0& 0& 0&$\frac{Z_{D_{s0}^*}^2 \sigma_y}{\sigma_y- \sigma_c}$& 0&$\frac{Z_{D_{s0}^*}^2 \sigma_c}{\sigma_c- \sigma_y}$\\
		$\sigma_{15} \ \sigma_{15}$& $\frac{1}{12}\sigma_x^2$& $\frac{1}{6}$& $\frac{1}{12}\sigma_y^2$& $\frac{1}{12}$& $\frac{3}{4}\sigma_c^2$&$\frac{3}{4}$\\
		$\sigma_8 \ \sigma_8$& $\frac{1}{6}\sigma_x^2$& $\frac{1}{3}$& $\frac{2}{3}\sigma_y^2$& $\frac{2}{3}$& 0&0\\
		$\sigma_0 \ \sigma_8$& $\frac{1}{2\sqrt{6}}\sigma_x^2$& $\frac{1}{\sqrt{6}}$& $\frac{1}{\sqrt{6}}\sigma_y^2$& $-\frac{1}{\sqrt{6}}$& 0&0\\
		$\sigma_0 \ \sigma_{15}$& $\frac{1}{4\sqrt{3}}\sigma_x^2$& $\frac{1}{2\sqrt{3}}$& $\frac{1}{4\sqrt{3}}\sigma_y^2$& $\frac{1}{4\sqrt{3}}$& $-\frac{\sqrt{3}}{4}\sigma_c^2$&$-\frac{\sqrt{3}}{4}$\\
		$\sigma_8 \ \sigma_{15}$& $\frac{1}{6\sqrt{2}}\sigma_x^2$& $-\frac{1}{3\sqrt{2}}$& $\frac{1}{3\sqrt{2}}\sigma_y^2$& $-\frac{1}{3\sqrt{2}}$& 0&0\\
		\hline\hline
		$ \pi_0 \ \pi_0$& 0& $\frac{1}{2}$& 0& $\frac{1}{4}$& 0&$\frac{1}{4}$\\
		$ \pi_1 \ \pi_1$& 0& $Z_\pi^2$& 0& 0& 0&0\\
		$\pi_4 \ \pi_4$& 0&  $\frac{Z_{K}^2\sigma_x}{\sigma_x +\sqrt{2} \sigma_y}$& 0& $\frac{Z_{K}^2\sqrt{2}\sigma_y}{\sqrt{2}\sigma_y +\sigma_x}$& 0&0\\
		$\pi_9 \ \pi_9$& 0& $\frac{Z_{D}^2\sigma_x}{\sigma_x+\sqrt{2} \sigma_c}$& 0& 0& 0& $\frac{Z_{D}^2\sqrt{2}\sigma_c}{\sigma_x+\sqrt{2} \sigma_c}$\\
		$\pi_{13} \ \pi_{13}$& 0& 0& 0& $\frac{Z_{D_{s}}^2 \sigma_y}{\sigma_y+ \sigma_c}$& 0&$\frac{Z_{D_{s}}^2 \sigma_c}{\sigma_y + \sigma_c}$\\
		$\pi_{15} \ \pi_{15}$& 0& $\frac{1}{6}$& 0& $\frac{1}{12}$& 0&$\frac{3}{4}$\\
		$\pi_8 \ \pi_8$& 0& $\frac{1}{3}$& 0& $\frac{2}{3}$& 0&0\\
		$\pi_0 \ \pi_8$& 0& $\frac{1}{\sqrt{6}}$& 0& $-\frac{1}{\sqrt{6}}$& 0&0\\
		$\pi_0 \ \pi_{15}$& 0& $\frac{1}{2\sqrt{3}}$& 0& $\frac{1}{4\sqrt{3}}$& 0&$-\frac{\sqrt{3}}{4}$\\
		$\pi_8 \ \pi_{15}$& 0& $\frac{1}{3\sqrt{2}}$& 0& $-\frac{1}{3\sqrt{2}}$& 0&0\\
		\hline\end{tabular}
	\caption{First and second derivatives of the squared quark mass with respect to the meson fields. }
	\label{tab:deriv}
\end{table}

\end{document}